\documentclass[12pt,preprint]{aastex}

\def\kms{\ifmmode{\rm km\th s^{-1}}\else km\th s$^{-1}$\fi}
\def\th{\thinspace}

\shortauthors{Torres et al.}
\shorttitle{HD~195987}

\begin{document}

\title{An interferometric-spectroscopic orbit for the binary
HD~195987: Testing models of stellar evolution for metal-poor
stars}

\author{Guillermo Torres\altaffilmark{1},
	Andrew F.\ Boden\altaffilmark{2,3},
	David W.\ Latham\altaffilmark{1},
	Margaret Pan\altaffilmark{1,4},
	Robert P.\ Stefanik\altaffilmark{1}
}

\email{gtorres@cfa.harvard.edu}

\email{*** To appear in The Astronomical Journal ***}
\email{*** September 2002 ***}

\altaffiltext{1}{Harvard-Smithsonian Center for Astrophysics, 60
Garden St., Cambridge MA 02138}

\altaffiltext{2}{Interferometry Science Center, California
Institute of Technology, 770 South Wilson Ave., Pasadena CA 91125}

\altaffiltext{3}{Department of Physics and Astronomy, Georgia State
University, 29 Peachtree Center Ave., Science Annex, Suite 400,
Atlanta GA 30303}

\altaffiltext{4}{Present address: Palomar Observatory, California
Institute of Technology, 770 South Wilson Ave., Pasadena CA 91125}

\begin{abstract}

We report spectroscopic and interferometric observations of the
moderately metal-poor double-lined binary system HD~195987, with an
orbital period of 57.3~days. By combining our radial-velocity and
visibility measurements we determine the orbital elements and derive
absolute masses for the components of $M_A = 0.844 \pm
0.018$~M$_{\sun}$ and $M_B = 0.6650 \pm 0.0079$~M$_{\sun}$, with
relative errors of 2\% and 1\%, respectively. We also determine the
orbital parallax, $\pi_{\rm orb} = 46.08 \pm 0.27$~mas, corresponding
to a distance of $21.70 \pm 0.13$~pc. The parallax and the measured
brightness difference between the stars in $V$, $H$, and $K$ yield the
component absolute magnitudes in those bands. We also estimate the
effective temperatures of the stars as $T_{\rm eff}^A = 5200 \pm
100$~K and $T_{\rm eff}^B = 4200 \pm 200$~K.  Together with detailed
chemical abundance analyses from the literature giving [Fe/H]$\approx
-0.5$ (corrected for binarity) and [$\alpha$/Fe]$=+0.36$, we use these
physical properties to test current models of stellar evolution for
metal-poor stars. Among the four that we considered, we find that no
single model fits all observed properties at the measured composition,
although we identify the assumptions in each one that account for the
discrepancy and we conclude that a model with the proper combination
of assumptions should be able to reproduce all the radiative
properties.  The indications from the isochrone fits and the pattern
of enhancement of the metals in HD~195987 are consistent with this
being a thick disk object, with an age of 10--12~Gyr. 

\end{abstract}


\section{Introduction}
\label{sec:introduction}

Accurate determinations of the physical properties of stars in binary
systems (mass, radius, temperature, etc.) provide for fundamental
tests of models of stellar structure and stellar evolution. The most
basic of those stellar properties is the mass. Several dozen eclipsing
binary systems have component mass and radius determinations that are
good to 1--2\% \citep[e.g.,][]{Andersen1991}, and show that
main-sequence models for stars with masses in the range from about
1~M$_{\sun}$ to 10~M$_{\sun}$ and heavy element abundances near solar
are in fairly good agreement with the observations. However, models
for stars with masses that are significantly higher or lower, or that
are in very early (pre-main sequence) or very advanced (post-main
sequence) stages of evolution, or models for chemical compositions
that are much different from solar, are largely untested by
observations due to a lack of suitable systems or a lack of accuracy. 

Stellar evolution theory has a wide range of applications in modern
astrophysics, some of which have profound cosmological implications.
One such application is the estimate of the ages of globular clusters,
which represent the oldest of the stellar populations in our Galaxy.
Because the metallicities of globular clusters are typically much
lower than solar, model fits to color-magnitude diagrams can be viewed
as useful tests of theory at these metal-poor compositions, at least
regarding the general shape of the isochrones. However, the predicted
mass at any given point along a fitted isochrone cannot be tested
directly against accurate observations because such a constraint is
unavailable for low metallicities.  While it is true that there is no
immediate cause for concern regarding the accuracy of current
metal-poor models, the use of such calculations for deriving the age
and other properties of globular clusters remains an extrapolation to
some degree, and it would be reassuring to have detailed observational
support. 

Double-lined spectroscopic binary systems with known compositions
below [Fe/H]$=-0.5$ that are also eclipsing and are therefore
particularly suitable for accurate determinations of the masses and
radii are virtually non-existent, both in clusters and in the general
field\footnote{A recent exception is the binary system OGLEGC-17 in
the globular cluster $\omega$~Cen, discovered by the OGLE microlensing
group \citep{Udalski1993}. A preliminary study by \citet{Thompson2001}
yielded encouraging results with errors in the masses and radii of
$\sim$7\% and $\sim$3\%, respectively, and a more detailed analysis
using additional observations can be expected in the near future
\citep[see][]{Kaluzny2001}. At an estimated metallicity of [Fe/H]$=
-2.29$ this is an extremely important system.}. But the absolute
masses can still be determined even when the binary is not eclipsing,
if the pair can be spatially resolved so that the inclination angle of
the orbit can be measured.  With the highly precise astrometry now
achievable using modern interferometers, this can significantly extend
the pool of objects available for testing models of stellar evolution
\citep[e.g.,][]{Armstrong1992,Hummel1994,Boden1999a,Boden2000,Hummel2001}. 

HD~195987 (HIP~101382, G209-35, Gliese~793.1, Groombridge~3215,
$\alpha = 20^h 32^m 51\fs6$, $\delta = +41\arcdeg 53\arcmin
55\arcsec$, J2000, $V=7.08$) is a binary with a period of 57.3~days
\citep{Imbert1980} reported to have a metallicity as low as
[m/H]$=-0.83$ \citep{Laird1988}, although more recent estimates are
not as low.  Originally listed as a single-lined spectroscopic binary,
further observations of HD~195987 have revealed the weak secondary
spectrum so that the system is now double-lined. The possibility of
eclipses has been mentioned occasionally in the literature, and would
obviously make the system even more interesting.  However, the
astrometric observations described in this paper clearly rule that
out.  The object was marginally resolved on one occasion by the
speckle technique \citep{Blazit1987}, but subsequent attempts have
failed to separate the components \citep{Balega1999,Mason2001a}.
Measurements by the HIPPARCOS mission \citep{Perryman1997} detected
the photocentric motion of the system with a semimajor axis of about
5~mas, and allowed the inclination angle and therefore the absolute
masses to be measured for the first time, albeit with relatively large
uncertainties \citep[see][]{Osborn1999}.  More recently the components
have been clearly resolved by the Palomar Testbed Interferometer
(PTI), opening the possibility of establishing the inclination angle
of the orbit with much higher accuracy.  This makes it a potentially
important system in which the absolute masses can be determined and
used to compare with stellar evolution theory for metal-poor stars,
and is the motivation for this paper. 

We report here our spectroscopic and interferometric observations of
HD~195987, the analysis of which has provided accurate masses (good to
1--2\%) and luminosities for the two components. We compare these
measurements with recent models for the appropriate metallicity. 
	
\section{Observations}
\label{sec:observations}

\subsection{Spectroscopy}
\label{subsec:spectroscopy}

HD~195987 was originally placed on the observing program at the
Harvard-Smithsonian Center for Astrophysics (CfA) as part of a project
to monitor the radial velocities of a sample of high proper motion
objects selected from the Lowell Proper Motion Survey
\citep{Giclas1971,Giclas1978}. The goal of this project is to
investigate a variety of issues related to Galactic structure
\citep[see][]{Carney1987}.  Observations were obtained from 1983 July
to 2001 January mostly with an echelle spectrograph mounted on the
1.5-m Wyeth reflector at the Oak Ridge Observatory (Harvard,
Massachusetts), and occasionally also with nearly identical
instruments on the 1.5-m Tillinghast reflector at the F.\ L.\ Whipple
Observatory (Mt.\ Hopkins, Arizona) and the Multiple Mirror Telescope
(also on Mt.\ Hopkins, Arizona) prior to its conversion to a
monolithic 6.5-m mirror. A single echelle order was recorded with
photon-counting intensified Reticon detectors at a central wavelength
of 5187~\AA, with a spectral coverage of 45~\AA.  The strongest
features present in this window are the lines of the \ion{Mg}{1}~b
triplet.  The resolving power is $\lambda/\Delta\lambda\approx
35,\!000$, and the signal-to-noise (S/N) ratios achieved range from
about 13 to 50 per resolution element of 8.5~\kms. A total of 73 usable
spectra were obtained over a period of more than 17 years. The
stability of the zero-point of our velocity system was monitored by
means of exposures of the dusk and dawn sky, and small systematic
run-to-run corrections were applied in the manner described by
\citet{Latham1992}. 

The observations for the first 3 years were analyzed by
\citet{Latham1988} with the techniques available at the time, but the
lines of the secondary are so faint that they escaped detection and
the object was treated as single-lined.  With the introduction of more
sophisticated analysis methods it was later discovered that the lines
from the secondary could in fact be seen \citep{Carney1994}.  A
preliminary analysis has been presented by \citet{Goldberg2002} on the
basis of the same spectra as the earlier study. For the present
investigation many more observations were obtained at higher S/N
ratios and at phases that contribute more information to the mass
determinations.  Radial velocities from all of the spectra were
derived by cross-correlation using the two-dimensional algorithm
TODCOR \citep{ZM94}.  The templates used for the primary and secondary
of HD~195987 were taken from an extensive library of synthetic spectra
based on model atmospheres by R.\ L.\ Kurucz\footnote{Available at
{\tt http://cfaku5.harvard.edu}.}, computed by Jon Morse specifically
for the wavelength window and resolution of our spectra. They are
available for a range of effective temperatures ($T_{\rm eff}$),
projected rotational velocities ($v \sin i$), surface gravities ($\log
g$), and metallicities ([m/H]).  Grids of correlations were run to
establish the template parameters giving the best match to each
component, based on the peak correlation value averaged over all
exposures and weighted by the S/N ratio of each spectrum.  The stars
present no measurable rotational broadening in our spectra, so we
adopted $v \sin i$ values of 0~\kms, along with $\log g$ values of
4.5, appropriate for dwarfs.  The temperature and metallicity
determinations are complicated by the fact that the secondary
component is extremely faint (only 10\% of the light of the primary;
see below), and in addition those two quantities are very strongly
correlated.  We fixed [m/H] at values ranging from $-1.0$ to $+0.5$
(in steps of 0.5~dex) and determined the effective temperatures in
each case. Lower metallicities lead to cooler temperatures because the
changes in the strength of the lines from these two effects tend to
compensate each other.  Formally, the best match to the observed
spectra was found for a metallicity around [m/H]$= -0.2$ and effective
temperatures of $T_{\rm eff}^A = 5350$~K and $T_{\rm eff}^B = 4550$~K
for the primary and secondary, respectively.  However, because of the
tradeoff mentioned above, different combinations of [m/H] and $T_{\rm
eff}$ give nearly equally good fits without changing the velocities
very much, and an external constraint is needed on one of these
parameters to break the degeneracy. For this we chose the metallicity,
since several determinations are available and appear to agree that
the system is metal-deficient compared to the Sun
(\S\ref{sec:physics}).  We adopted the value [m/H]$ = -0.5$, which
leads to effective temperatures of $T_{\rm eff}^A = 5200 \pm 150$~K
and $T_{\rm eff}^B = 4450 \pm 250$~K.  The light ratio in our spectral
window was found to be $(l_B/l_A)_{5187} = 0.09 \pm 0.01$. 

Systematic errors in the radial velocities resulting from the narrow
spectral window were investigated by means of numerical simulations,
as discussed in detail by \citet{Latham1996}.  Briefly, we generated a
set of artificial binary spectra by combining the primary and
secondary templates in the appropriate ratio and applying velocity
shifts for both components as computed from a preliminary orbital
solution at the actual times of observation of each of our spectra.
These artificial spectra were then processed with TODCOR in exactly
the same way as the real observations, and the resulting velocities
were compared with the input (synthetic) values.  The differences were
found to be smaller than 0.4~\kms, but were nevertheless applied to the
raw velocities as corrections. The effect on the minimum masses
derived from the spectroscopic orbit, which depend on the velocity
semi-amplitudes $K_A$ and $K_B$, is at the level of 0.5\%. The final
velocities, including corrections, are listed in Table~\ref{tab:rvs}. 


The spectroscopic orbital elements resulting from these velocities are
given in column~2 of Table~\ref{tab:sb2}. They are the period ($P$, in
days), the center-of-mass velocity ($\gamma$, in \kms), the radial
velocity semi-amplitudes of the primary and secondary ($K_A$ and
$K_B$, in \kms), the orbital eccentricity ($e$), the longitude of
periastron of the primary ($\omega_A$, degrees), and the time of
periastron passage ($T$, Julian days).  Earlier single-lined orbital
solutions for HD~195987 were published by \citet{Imbert1980},
\citet{Latham1988}, and \citet{Duquennoy1991}, and
\citet{Goldberg2002} recently reported a double-lined solution. They
are included in Table~\ref{tab:sb2} for comparison with ours. The
orbit published by \citet{Latham1988}, which is based on a small
subset of the same spectra used in the present paper, was superseded
by that by \citet{Goldberg2002} that used the same material, which in
turn is superseded by our new definitive results.  The solution by
\citet{Duquennoy1991} is not independent of that by
\citet{Imbert1980}, but represents an update using additional
observations with the same instrument.  As seen in the table, all
these solutions are fairly similar. 


The uncertainties given in Table~\ref{tab:sb2} for the velocity
amplitudes of our new solution, upon which the masses of the
components depend critically, are strictly internal errors.  In
addition to the biases described earlier, further systematic errors in
the velocities can occur because of uncertainties in the template
parameters, particularly the temperatures and the metallicity.
Extensive grids of correlations showed that the sensitivity of the
velocity amplitudes to the secondary temperature is minimal because
that star is so faint. They also showed that the combined effects of
errors in the primary temperature and the metallicity contribute an
additional $0.04~\kms$ uncertainty to $K_A$ and $0.3~\kms$ to $K_B$.
These have been combined with the internal errors in the amplitudes,
and propagated through to the final masses that we report later in
\S\ref{sec:physics}. 
	
\subsection{Interferometry}
\label{sec:interferometry}

Near-infrared, long-baseline interferometric measurements of HD~195987
were conducted with the Palomar Testbed Interferometer (PTI), which is
a 110-m baseline $H$- and $K$-band ($\lambda\sim$1.6~$\mu$m and
$\sim$2.2~$\mu$m) interferometer located at Palomar Observatory. It is
described in full detail elsewhere \citep{Colavita1999a}. The
instrument gives a minimum fringe spacing of about 4~mas at the sky
position of our target, making the HD~195987 binary system readily
resolvable. 
	
The interferometric observable used for these measurements is the
normalized fringe contrast or {\em visibility} (squared), $V^2$, of an
observed brightness distribution on the sky.  The analysis of such
data in the context of a binary system is discussed in detail by
\citet{Hummel1998}, \citet{Boden1999a}, \citet{Boden1999b},
\citet{Boden2000}, and \citet{Hummel2001}, and will not be repeated
here. 

HD~195987 was observed with PTI in conjunction with objects in our
calibrator list in the $K$ band ($\lambda\sim$2.2~$\mu$m) on 32 nights
between 1999 June 24 and 2001 September 28, covering roughly 14
periods of the system.  Additionally, the target was observed in the
$H$ band ($\lambda\sim$1.6~$\mu$m) on five nights between 2000 July 20
and 2001 September 27.  HD~195987, along with the calibration objects,
was observed multiple times during each of these nights, and each
observation, or scan, was approximately 130 seconds long.  For each
scan we computed a mean $V^2$ value from the scan data, and the error
in the $V^2$ estimate from the rms internal scatter
\citep{Colavita1999b}.  HD~195987 was always observed in combination
with one or more calibration sources within $\sim$20$\arcdeg$ on the
sky.  For our study we used four stars as calibrators: HD~195194,
HD~200031, HD~177196, and HD~185395.  Table~\ref{tab:calibrators}
lists the relevant physical parameters for these objects. The last two
calibration objects are known to have at least one visual companion
\citep[Washington Double Star Catalog;][]{Mason2001b} that could
conceivable affect our visibility measurements.  The companion of
HD~177196 is so distant (44\arcsec\ at last measurement in 1925) that
it has no effect because of the effective 1\arcsec\ field stop of the
PTI fringe camera focal plane \citep{Colavita1999a}. Of the three
recorded companions of HD~185395 two are more than 40\arcsec\ away,
and the other has an angular separation of 2\farcs9. This close
companion is more than 5 magnitudes fainter than the primary in the
$K$ band, and can therefore also be safely ignored. 
	

The calibration of the HD~195987 $V^2$ data was performed by
estimating the interferometer system visibility ($V^{2}_{\rm sys}$)
using calibration sources with model angular diameters, and then
normalizing the raw HD~195987 visibility by $V^{2}_{\rm sys}$ to
estimate the $V^2$ measured by an ideal interferometer at that epoch
\citep{Mozurkewich1991,Boden1998}.  Calibrating our HD~195987 dataset
with respect to the four calibration objects listed in
Table~\ref{tab:calibrators} results in a total of 171 calibrated scans
(134 in $K$, 37 in $H$) on HD~195987 on 37 different nights.  Our
calibrated synthetic wide-band visibility measurements in the $H$ and
$K$ bands are summarized in Table~\ref{tab:visibH} and
Table~\ref{tab:visibK}, which include the time of observation, the
calibrated $V^2$ measurement and its associated error, the residual
from the final fit (see below), the ($u$,$v$) coordinates in units of
the wavelength $\lambda$ (weighted by the S/N ratio), and the orbital
phase for each of our observations. 



\section{Determination of the orbit}
\label{sec:orbit}

The radial velocities and interferometric visibilities for HD~195987
contain complementary information on the orbit of the system, which is
described by the 7 elements mentioned in \S\ref{subsec:spectroscopy}
and the additional elements $a$ (relative semimajor axis, expressed
here in mas), $i$ (inclination angle), and $\Omega$ (position angle of
the ascending node).  Only by combining data from both techniques can
all 10 elements of the three-dimensional orbit be determined. While
the center-of-mass velocity and the velocity amplitudes depend only on
the spectroscopy, the information on $a$, $i$, and $\Omega$ is
contained solely in the interferometric visibilities. However, the
$V^2$ values are invariant to a $180\arcdeg$ change in the position
angle of the binary, so distinguishing between the descending node and
the ascending node (where, according to convention, the secondary is
receding) relies on our knowledge of the radial velocities. 

As in previous analyses using PTI data
\citep{Boden1999a,Boden1999b,Boden2000,BL2001}, the orbit of HD~195987
is solved by fitting a Keplerian orbit model directly to the
calibrated $V^2$ and radial velocity data simultaneously.  This allows
the most complete and efficient use of the observations\footnote{In
principle it would be possible to derive the position angle ($\theta$)
and the angular separation ($\rho$) of the binary from the visibility
measurements for each night.  These intermediate data could then be
used instead of the visibilities in what might be considered a more
conventional astrometric-spectroscopic solution. However, the result
would be inferior to the fit we describe here for several reasons
\citep[see, e.g.,][]{Hummel2001}, not the least of which is the fact
that, due to the pattern of our observations, the number of scans on
many nights is insufficient to solve for the separation vectors, which
would result in a rather serious loss of information (phase coverage).
Therefore, we refrain from even listing values of \{$\rho$, $\theta$\}
in Table~\ref{tab:visibH} and Table~\ref{tab:visibK}.}, so long as the
two types of measurements are free from systematic errors (see below).
It also takes advantage of the redundancy between spectroscopy and
interferometry for the four elements $P$, $e$, $\omega_A$, and $T$.
Relative weights for the velocities (which are different for the
primary and the secondary) and the visibilities were applied based on
the internal errors of each type of observation. 

Formally the interferometric visibility observables have the potential
for resolving not only the HD~195987 relative orbit, but the binary
components themselves, and these two effects must be considered
simultaneously \citep[e.g.,][]{Hummel1994}.  However, in the case of
HD~195987 at a distance of approximately 22~pc the dwarf components
have typical apparent sizes of less than 0.5~mas, which are highly
unresolved by the 3--4~mas PTI fringe spacing.  We have therefore
estimated the apparent sizes of the components using the bolometric
flux and effective temperatures \citep[see][and references
therein]{Blackwell94}, and constrained the orbital solutions to these
model values (\S\ref{sec:physics}).  We have adopted apparent
component diameters of $0.419 \pm 0.018$~mas and $0.314 \pm 0.049$~mas
for the primary and secondary components, respectively.  These values
are much smaller than the PTI $H$ and $K$ fringe spacings and have a
negligible effect on the parameters of the orbit. 

The results of this joint fit are listed in Table~\ref{tab:combsol}
(``Full-fit" solution), where we have used the $H$-band and $K$-band
visibilities simultaneously, weighted appropriately by the
corresponding errors.  In addition to the orbital elements, the
visibility measurements provide the intensity ratio in $H$ and $K$,
which are critical for the model comparisons described later. 


The uncertainties listed for the elements in Table~\ref{tab:combsol}
include both a statistical component (measurement error) and our best
estimate of the contribution from systematic errors, of which the main
components are: (1) uncertainties in the calibrator angular diameters
(Table~\ref{tab:calibrators}); (2) the uncertainty in the center-band
operating wavelength ($\lambda_0 \approx$ 1.6~$\mu$m and 2.2~$\mu$m),
taken to be 20~nm ($\sim$1\%); (3) the geometrical uncertainty in our
interferometric baseline ($< 0.01$\%); and (4) listed uncertainties in
the angular diameters assumed for the stars in HD~195987, that were
held constant in the fitting procedure. 

\begin{figure} 
\epsscale{0.8} 
\plotone{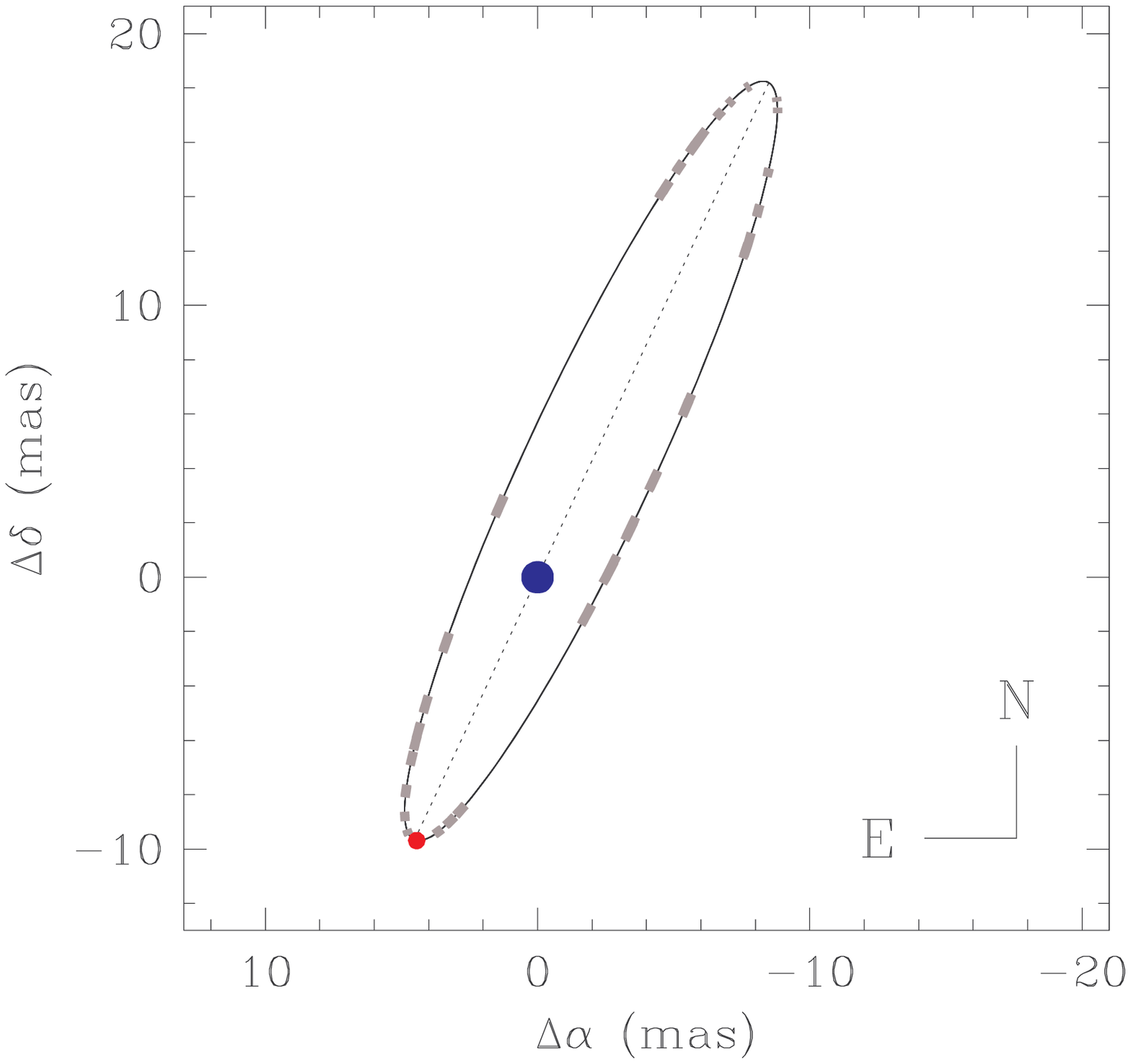}
 \figcaption[Torres.fig01.ps]{Relative orbit of HD~195987 on the plane
of the sky, with the primary at the origin and the secondary shown at
periastron.  Motion is clockwise, and the dotted line indicates the
line of nodes.  The heavy line segments along the relative orbit
indicate areas where we have phase coverage in our $H$- and $K$-band
PTI data (they are not separation vector estimates). Our data sample
essentially all phases of the orbit well, leading to a reliable orbit
determination.  Component diameters are rendered to a scale three
times larger than their actual size in relation to the
orbit.\label{fig:hd195987_orbit}} 
 \end{figure}

\placefigure{fig:hd195987_orbit}

\begin{figure}
\epsscale{1.0}
\plotone{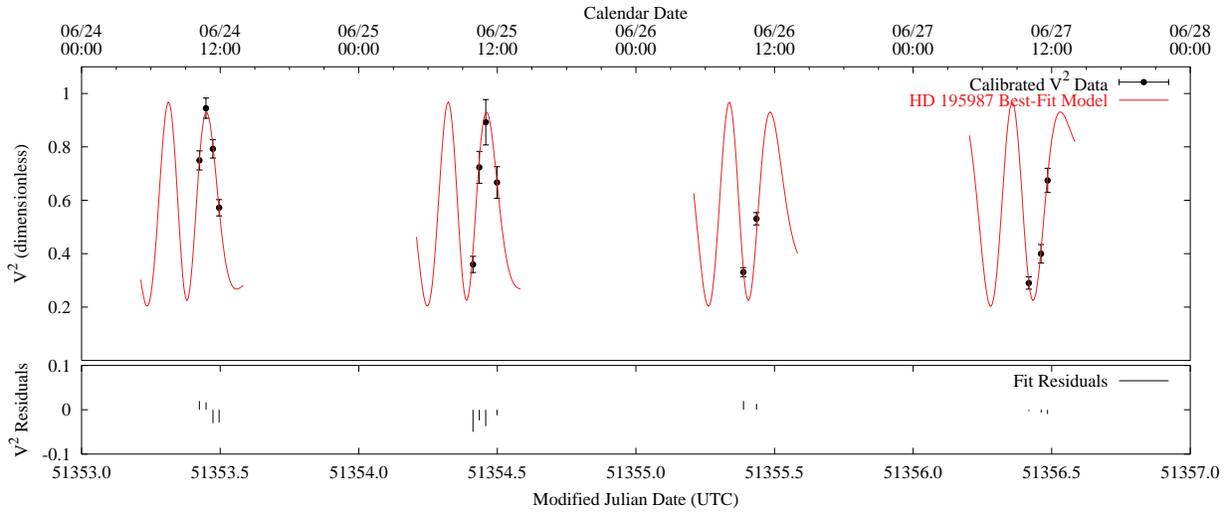}
 \figcaption[Torres.fig02.ps]{$V^2$ model for HD~195987.  The top
panel shows a direct comparison between four consecutive nights
(24--27 June 1999) of calibrated $K$-band $V^2$ data and the
corresponding predictions computed from our ``Full-fit'' orbit model
for HD~195987 (Table~\ref{tab:combsol}).  The lower frame gives the
residuals between the data and the model. The modified Julian Date is
MJD = HJD $-$ 0.5.  \label{fig:hd195987_V2trace}}
 \end{figure}

\begin{figure}
\epsscale{1.0}
\plotone{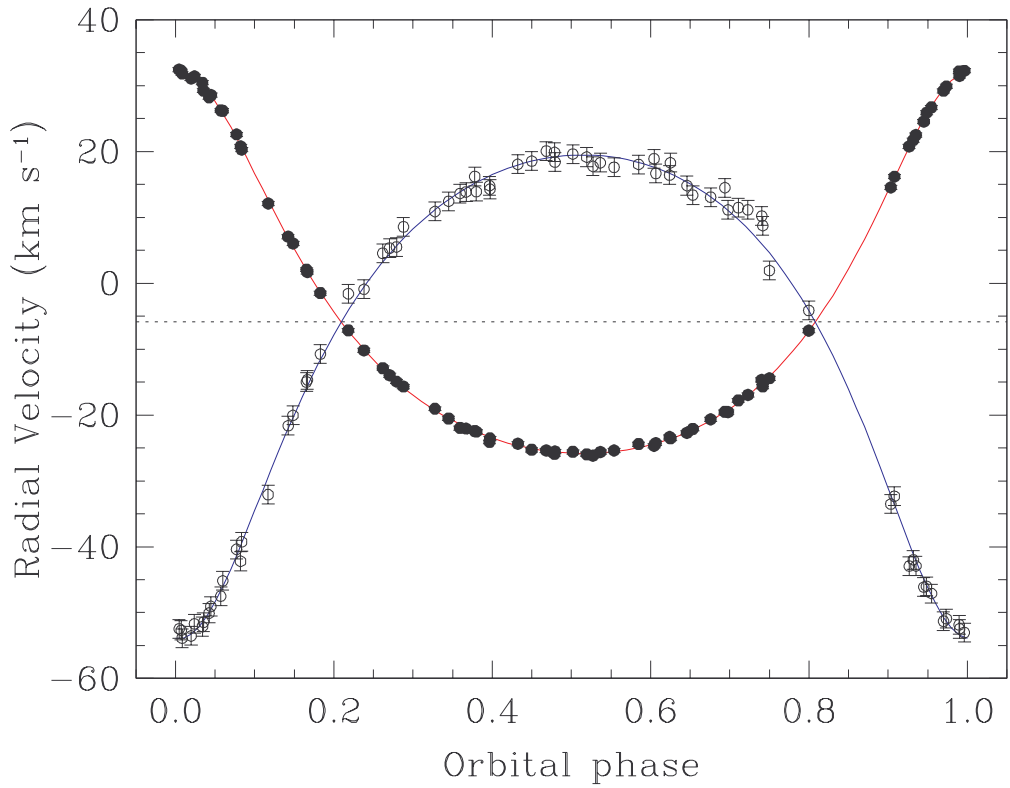}
 \figcaption[Torres.fig03.ps]{Radial velocity observations and
spectroscopic orbit of HD~195987 from our `Full-fit" solution
(Table~\ref{tab:combsol}).  The primary is represented with filled
circles, and the center-of-mass velocity is indicated by the dotted
line. The error bars for the primary are smaller than the point
size.\label{fig:hd195987_RVfit}}
 \end{figure}

\begin{figure}
\epsscale{1.0}
\vskip -1.0in
\plotone{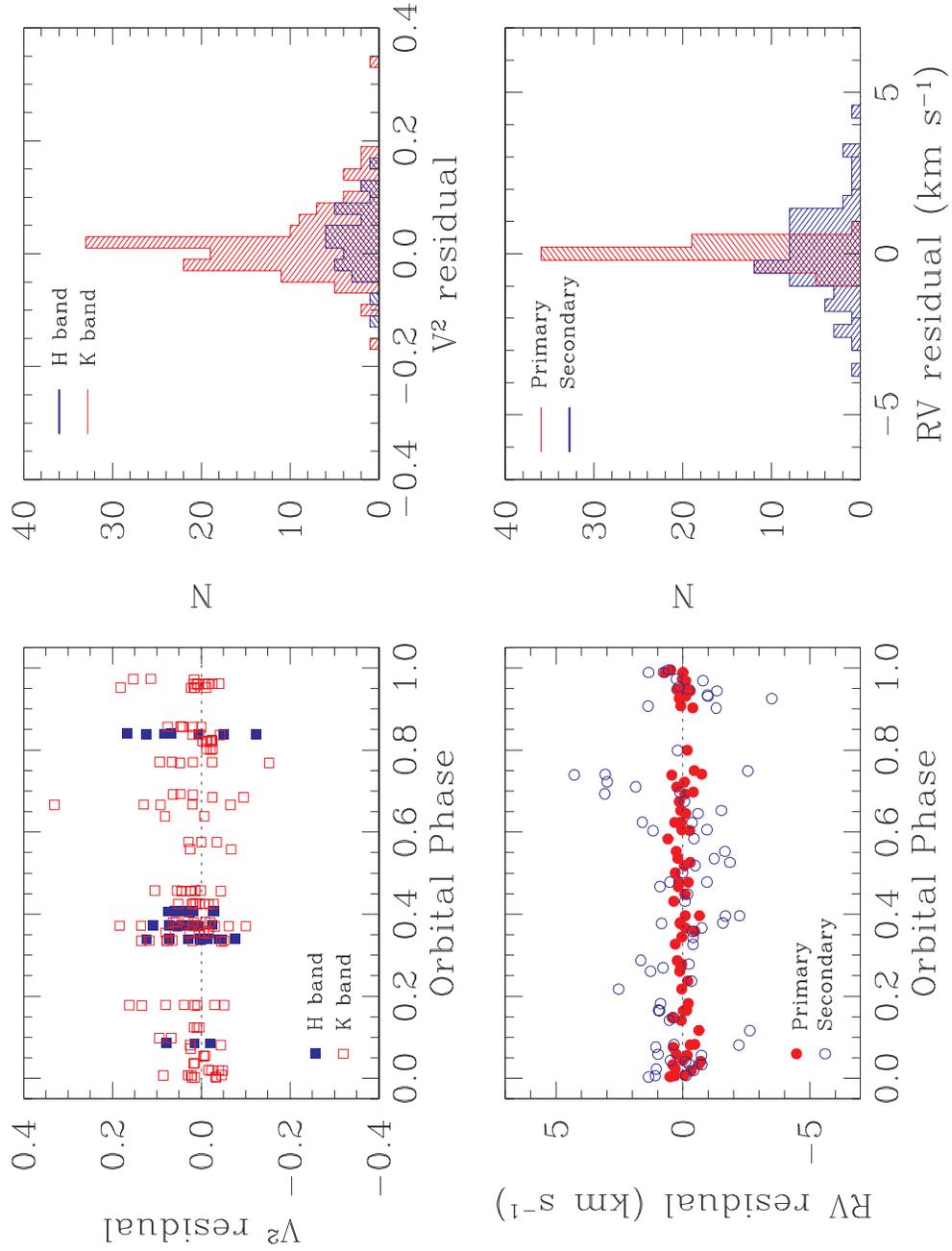}
\vskip -0.8in
 \figcaption[Torres.fig04.ps]{Residuals from the orbital fit for
HD~195987 (Table~\ref{tab:combsol}) as a function of orbital phase,
for the $H$- and $K$-band visibilities ($V^2$) and for the radial
velocities, along with the corresponding
histograms.\label{fig:hd195987_fitResiduals}}
 \end{figure}

Figure~\ref{fig:hd195987_orbit} depicts the relative visual orbit of
the HD~195987 system on the plane of the sky, with the primary
component rendered at the origin and the secondary component shown at
periastron.  We have indicated the phase coverage of our $V^2$ data on
the relative orbit with heavy line segments. Our interferometric data
sample essentially all phases of the orbit (see also
Figure~\ref{fig:hd195987_fitResiduals}), leading to a reliable
determination of the elements. The fit to these data is illustrated in
Figure~\ref{fig:hd195987_V2trace}, which shows four consecutive nights
of PTI visibility data on HD~195987 (24--27 June 1999), and $V^2$
predictions based on the ``Full Fit'' model for the system
(Table~\ref{tab:combsol}). The fit to the radial velocity measurements
is shown in Figure~\ref{fig:hd195987_RVfit}. Phase plots of the $V^2$
and velocity residuals are shown in
Figure~\ref{fig:hd195987_fitResiduals}, along with the corresponding
histograms. We note here in passing that the agreement between the
speckle measurement by \citet{Blazit1987} (which by chance was
obtained very near periastron passage, at phase 0.981) and our
interferometric-spectroscopic orbit is very poor. However, this is not
surprising given the limited resolution of the speckle observation
(32~mas in the visible) compared to the predicted separation at the
time of the observation (10.5~mas). 

\placefigure{fig:hd195987_V2trace}
\placefigure{fig:hd195987_RVfit}

For comparison, we list also in Table~\ref{tab:combsol} the solutions
we obtain for HD~195987 using only the interferometric visibilities
(``$V^2$-only"), and only the radial velocities (``RV-only", repeated
from Table~\ref{tab:sb2}). These two separate fits give rather similar
results, indicating no significant systematic differences between the
spectroscopic and interferometric data sets. Also included in the
table is the solution reported in the HIPPARCOS Catalogue, where the
semimajor axis refers to the photocentric motion of the pair rather
than the relative motion. In the HIPPARCOS solution the elements $P$,
$e$, $\omega_A$, and $T$ were adopted from the work by
\citet{Duquennoy1991}, and held fixed. Given the much larger formal
errors, the resulting elements are as consistent with our solution as
can be expected. 

\placefigure{fig:hd195987_fitResiduals}

\section{The light ratio in the optical}
\label{sec:lightratio}

The interferometric measurements of HD~195987 with PTI provide the
intensity ratio between the components in the $H$ and $K$ bands, and
allow the individual luminosities to be determined in the infrared. A
similar determination in the optical based on our spectroscopy was
given in \S\ref{subsec:spectroscopy}. A small correction from the
5187~\AA\ region to the visual band yields $(l_B/l_A)_V = 0.10 \pm
0.02$. Because the secondary is so faint, this estimate may be subject
to systematic errors that are difficult to quantify. 

The detection of the photocentric motion of the binary by HIPPARCOS,
along with the measurement of the apparent separation with PTI, allows
an independent estimate of the light ratio to be made in the optical.
The relative semimajor axis ($a$) and the photocentric semimajor axis
($\alpha$) are related by the classical expression $\alpha = a (B -
\beta)$, where $B = M_B/(M_A+M_B)$ is the mass fraction (also
expressed as $B=K_A/(K_A+K_B)$ in terms of the observables) and
$\beta=l_B/(l_A+l_B)$ is the fractional luminosity.  The ratio
$l_B/l_A$ at the effective wavelength of the HIPPARCOS observations
($H_p$ passband) can therefore be determined. 

In order to take advantage of the much improved orbital elements
compared to those available to the HIPPARCOS team, we have re-analyzed
the HIPPARCOS intermediate astrometric data (abscissae residuals) as
described in the original Catalogue \citep{HIP1997} \citep[see
also][]{Pourbaix2000}.  The orbital period, the eccentricity, the
longitude of periastron, and the time of periastron passage were
adopted from our combined visibility-radial velocity fit as listed in
Table~\ref{tab:combsol}, and held fixed. In addition, the inclination
angle and the position angle of the node are now known much better
than can be determined from the HIPPARCOS data, so we adopted our own
values here as well. The only parameters left to determine are then
the semimajor axis of the photocenter ($\alpha$), the corrections to
the position and proper motion of the photocenter, and the correction
to the HIPPARCOS parallax. 

The relevant results from this analysis are $\alpha = 5.04 \pm
0.47$~mas and $\pi_{\rm HIP} = 45.30 \pm 0.46$~mas. The uncertainties
may be somewhat underestimated because they do not account for
correlations between the elements (since we have fixed several of them
in the new solution), although we do not expect the effect to be large
in this case.  As it turns out, the new results are not very different
from the original determinations by the HIPPARCOS team ($\alpha = 5.24
\pm 0.66$~mas and $\pi_{\rm HIP} = 44.99 \pm 0.64$~mas). Forcing the
HIPPARCOS parallax to come out identical to our orbital parallax makes
a negligible change in the semimajor axis of the photocenter. 

The luminosity ratio we determine from this is $(l_B/l_A)_V = 0.13 \pm
0.04$, which includes a very small correction from the $H_p$ band to
the Johnson $V$ band.  This is in good agreement with our
spectroscopic determination, and we adopt the weighted average of the
two results, $(l_B/l_A)_V = 0.11 \pm 0.02$, which corresponds to
$\Delta V = 2.4 \pm 0.2$~mag. 
	
\section{Physical properties of HD~195987}
\label{sec:physics}

The absolute masses of the components follow from the results of our
combined interf\-erometric-spectroscopic solution summarized in
Table~\ref{tab:combsol}, and are $M_A = 0.844 \pm 0.018$~M$_{\sun}$
and $M_B = 0.6650 \pm 0.0079$~M$_{\sun}$\footnote{These uncertainties
include the contribution due to the template parameters discussed in
\S\ref{subsec:spectroscopy}.}. The high accuracy achieved (2.1\% error
for the primary and 1.2\% error for the secondary) is the result of
the high quality of the observations as well as the favorable geometry
of the system ($\sin i \approx 0.99$). The limiting factor is the
spectroscopy, and in particular the velocities for the faint
secondary. 

The combination of the angular separation from interferometry and the
linear separation from spectroscopy yields the orbital parallax of the
system independently of any assumptions beyond Newtonian physics. The
result, $\pi_{\rm orb} = 46.08 \pm 0.27$~mas (distance$= 21.70 \pm
0.13$~pc) is nearly a factor of two more precise than our revised
HIPPARCOS value of $\pi_{\rm HIP} = 45.30 \pm 0.46$~mas, and the two
determinations differ by about 1.5$\sigma$ (1.7\%). The corresponding
difference in the distance modulus, and consequently in the absolute
magnitudes derived below, is $\Delta(m-M) = 0.037$~mag. The
ground-based trigonometric parallax of HD~195987 from the weighted
average of 5 determinations is $\pi = 49.1 \pm 5.1$~mas
\citep{vanaltena1995}. 

The component luminosities in the $V$, $H$, and $K$ passbands depend
on the flux ratios that we have measured, on the orbital parallax, and
on the combined-light photometry.  For the $V$ magnitude of the system
we adopt the average value given by \citet{Mermilliod1994}, which is
the result of 10 individual measurements from 7 different sources: $V
= 7.080 \pm 0.016$.  Two measurements of $K$ are available from
\citet{Voelcker1975} and from the 2MASS Catalog \citep{2MASS}.
Conversion of both to the CIT system \citep{Elias1983}, using the
transformations by \citet{Bessell1988} and \citet{Carpenter2001},
gives $K=4.98 \pm 0.03$.  The $H$ magnitude of the system measured by
\citet{Voelcker1975} is $H = 5.23 \pm 0.05$. A 2MASS measurement in
$H$ is not available because the object is saturated in this passband,
although not in $J$ and $K$.  This is somewhat unexpected if we assume
normal colors for stars of this mass, and it casts some doubt on the
reliability of the $H$-band measurement.  As a check we obtained a new
measurement using an infrared camera on the 1.2-m telescope at the F.\
L.\ Whipple Observatory equipped with a $256\times 256$ InSb detector
array, with standard stars adopted from \citet{Elias1982} and 2MASS.
The result, $H = 5.04 \pm 0.03$ (on the CIT system), is significantly
different from the previous estimate, confirming our suspicions, and
it is the value we adopt for the remainder of the paper.  Any
extinction and reddening corrections are negligible at a distance of
only 22~pc.  The individual absolute magnitudes and colors are listed
in Table~\ref{tab:physics} along with the other physical properties. 


The $V\!-\!K$ color indices for each component allow for an
independent estimate of the effective temperatures of the stars.
Based on the color-temperature calibrations by \citet{Martinez1992},
\citet{Alonso1996}, and \citet{Carney1994} (the later being for the
CIT system), we obtain average values of $T_{\rm eff}^A = 5200 \pm
100$~K and $T_{\rm eff}^B = 4100 \pm 200$~K that are nearly
independent of the adopted metal abundance (for which these
calibrations include a corrective term). The primary estimate is
identical to our spectroscopic value in \S\ref{subsec:spectroscopy},
while the secondary estimate is somewhat lower than the spectroscopic
value and its uncertainty has a substantial contribution from the
photometric errors. 

An additional estimate of the effective temperature may be obtained
from photometry in other bands, via deconvolution of the combined
light. For this we used the indices for HD~195987 in the Johnson and
Str\"omgren systems as listed by \citet{Mermilliod1994} ($B\!-\!V =
0.800 \pm 0.014$; $U\!-\!B = 0.370 \pm 0.026$) and by
\citet{Olsen1983,Olsen1993} ($b\!-\!y = 0.480 \pm 0.002$, $m_1 = 0.296
\pm 0.005$, $c_1 = 0.271\pm 0.005$).  The deconvolution was performed
using tables of standard colors for normal stars by
\citet{Lejeune1998} (for a metallicity of [Fe/H]$= -0.5$) and by
\citet{Olsen1984}, and adopting a magnitude difference between the
primary and secondary in the visual band of $\Delta V = 2.4$~mag
(\S\ref{sec:lightratio}). In addition to the calibrations mentioned
above we used those by \citet{Carney1983} and \citet{Olsen1984}.  The
results based on the deconvolved $B\!-\!V$, $b\!-\!y$, and also the
$V\!-\!K$ indices for the system (accounting for the small difference
between the different photometric systems in the infrared) give values
of $5200 \pm 100$~K for the primary and $4100 \pm 200$~K for the
secondary, identical to the estimates from the observed $V\!-\!K$
colors. 

We adopt for the stars in HD~195987 the weighted average of our three
determinations of the effective temperature (one spectroscopic and two
photometric): $T_{\rm eff}^A = 5200 \pm 100$~K and $T_{\rm eff}^B =
4200 \pm 200$~K.  From these effective temperatures, an estimated
system bolometric flux of ($5.308\pm 0.062$) $\times$ 10$^{-8}$ erg
cm$^{-2}$ s$^{-1}$ based on archival broadband photometry, and our
observed 2.2~$\mu$m intensity ratio we estimate component bolometric
fluxes of ($4.28\pm 0.19$) and ($1.02\pm 0.28$) $\times$ 10$^{-8}$ erg
cm$^{-2}$ s$^{-1}$.  These combined again with the adopted component
temperatures yield estimated apparent diameters of $0.419\pm0.018$~mas
and $0.314\pm0.049$~mas for the primary and secondary components,
respectively, which are the values adopted for the orbital fit
(\S\ref{sec:orbit}).  At our estimated system distance to HD~195987
these angular diameters imply physical radii of
$0.979\pm0.043$~R$_{\sun}$ and $0.73\pm0.11$~R$_{\sun}$ for the
primary and secondary components, respectively. 

As mentioned in \S\ref{sec:introduction}, early high-resolution
spectroscopic estimates of the metal abundance of HD~195987 placed the
system in an interesting range for absolute mass determinations, at a
metallicity of [m/H]$ = -0.83 \pm 0.15$ \citep{Laird1988}.
Subsequently the same authors revised their estimate upward to [m/H]$
= -0.60 \pm 0.15$ using the same techniques and more spectra
\citep{Carney1994}.  Other determinations in the literature have given
values that are similar or somewhat closer to the solar abundance.
\citet{Fulbright2000} used high-resolution spectroscopy to derive an
estimate of [Fe/H]$ = -0.66 \pm 0.13$, while \citet{Beers1999}
obtained [Fe/H]$ = -0.52 \pm 0.20$ with a combination of different
spectroscopic methods based on lower-resolution spectra.
\citet{Marsakov1988} inferred [Fe/H]$ = -0.34$ based on the
ultraviolet excess of the system, and \citet{Wyse1995} used
Str\"omgren photometry to estimate the metallicity at [Fe/H]$ =
-0.31$. Despite the range of values, these sources do seem to point
toward a heavy element abundance moderately lower than the Sun, by
perhaps a factor of 3 or so. We discuss possible biases in the
metallicity estimates in \S\ref{sec:iron}. 

\section{Discussion}
\label{sec:discussion}

With many of the fundamental physical properties of HD~195987 now
known to high accuracy, we proceed in this section with a detailed
comparison between the observations and recent models of stellar
evolution. In addition to the prospect of an interesting comparison
for a chemical composition lower than solar, both components of our
system have masses below 1~M$_{\sun}$, a regime in which relatively
few stars have absolute mass determinations good to 1--2\% \citep[see,
e.g.,][]{Andersen1991,Clausen1999,Delfosse2000}. 

\subsection{Comparison with stellar evolution models}
\label{sec:modelcomp}

In order for a given model to be successful, we require that it agree
with all the measurements for HD~195987 \emph{simultaneously}, and
\emph{for both components} at the same time since they are presumably
coeval.  The measurements are the absolute masses, the absolute
magnitudes in three different passbands ($M_V$, $M_H$, $M_K$), the
effective temperatures, and the metal abundance. 

\begin{figure}
\epsscale{1.15}
\plotone{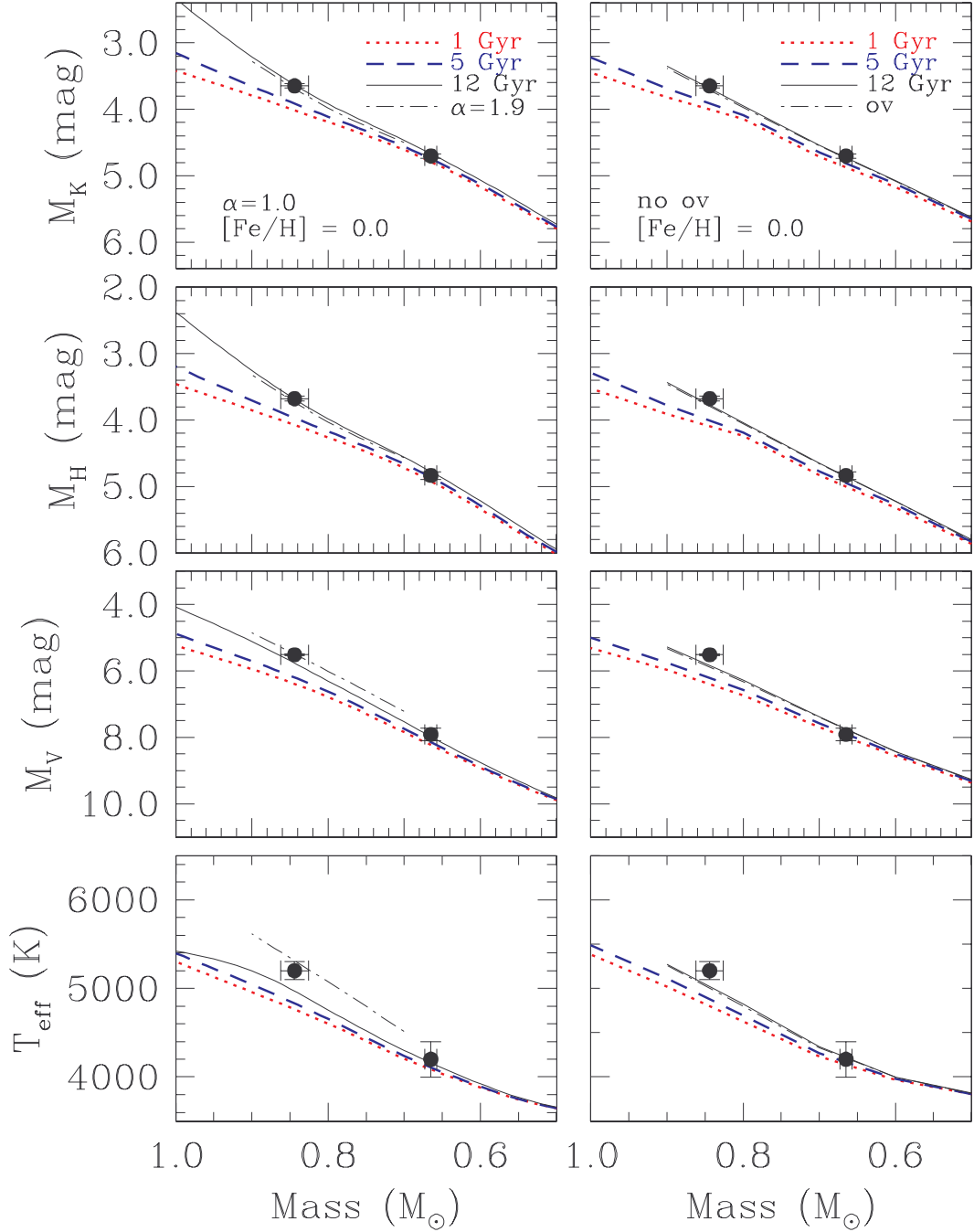}
 \figcaption[Torres.fig05.ps]{Comparison between the observations for
HD~195987 and the Yale-Yonsei models by Yi et al.\ (2001).  The left
panels show isochrones for different ages, as labeled, and the right
panels show the effect of a change in the metallicity. The infrared
magnitudes from the original isochrones have been transformed to the
CIT system \citep{Bessell1988}.\label{fig:yale}}
 \end{figure}

In Figure~\ref{fig:yale} we show the absolute magnitudes and
temperatures of the primary and secondary of HD~195987 as a function
of mass, compared with isochrones from the Yale-Yonsei models by
\citet{Yi2001}\footnote{The infrared magnitudes in these isochrones
are based on the color tables by \citet{Lejeune1998}, which adopt
filter transmission functions from \citet{Bessell1988} for the
Johnson-Glass system. For the comparison with our observations we
transformed the isochrones to the CIT system using the corrections by
\citet{Bessell1988}.}.  Changes in the isochrones with age for a fixed
metallicity equal to solar are illustrated on the left, and consist
not only of a vertical displacement but also of a change in slope,
becoming steeper for older ages. A similar effect is seen in the
diagrams on the right-hand side, which show changes in the models for
a fixed age (12~Gyr) and a range of metallicities bracketing the
values reported for HD~195987.  The isochrones shown with a solid
line, which correspond to solar metallicity and an age of 12~Gyr,
provide a fairly good fit to the absolute magnitudes in the $H$ and
$K$ bands, as well as to the $M_V$ of the primary star. However, the
secondary component appears underluminous in $V$, or else the slope of
the models in the mass-$M_V$ plane is too shallow.  In addition, the
temperatures predicted by the models may be somewhat too hot,
particularly for the secondary.  Setting these differences aside for
the moment, the \citet{Yi2001} models would appear to point toward a
metallicity for HD~195987 that is not far from solar, along with a
fairly old age.  This metallicity seems to conflict with the
observational evidence indicated earlier.  Formally the best fit to
the magnitudes in $H$, $K$, and $V$ (primary only) gives a metallicity
of [Fe/H]$ = -0.07$ and an age of 11.5~Gyr.  Lowering the age of the
isochrones to 10~Gyr and at the same time decreasing the metallicity
to [Fe/H]$ = -0.15$ still produces tolerably good fits, but
metallicities much lower than this are inconsistent with the measured
magnitudes for any age.  The disagreement with the $M_V$ of the
secondary and with the temperatures remains. 

\placefigure{fig:yale}

\begin{figure}
\epsscale{1.1}
\plotone{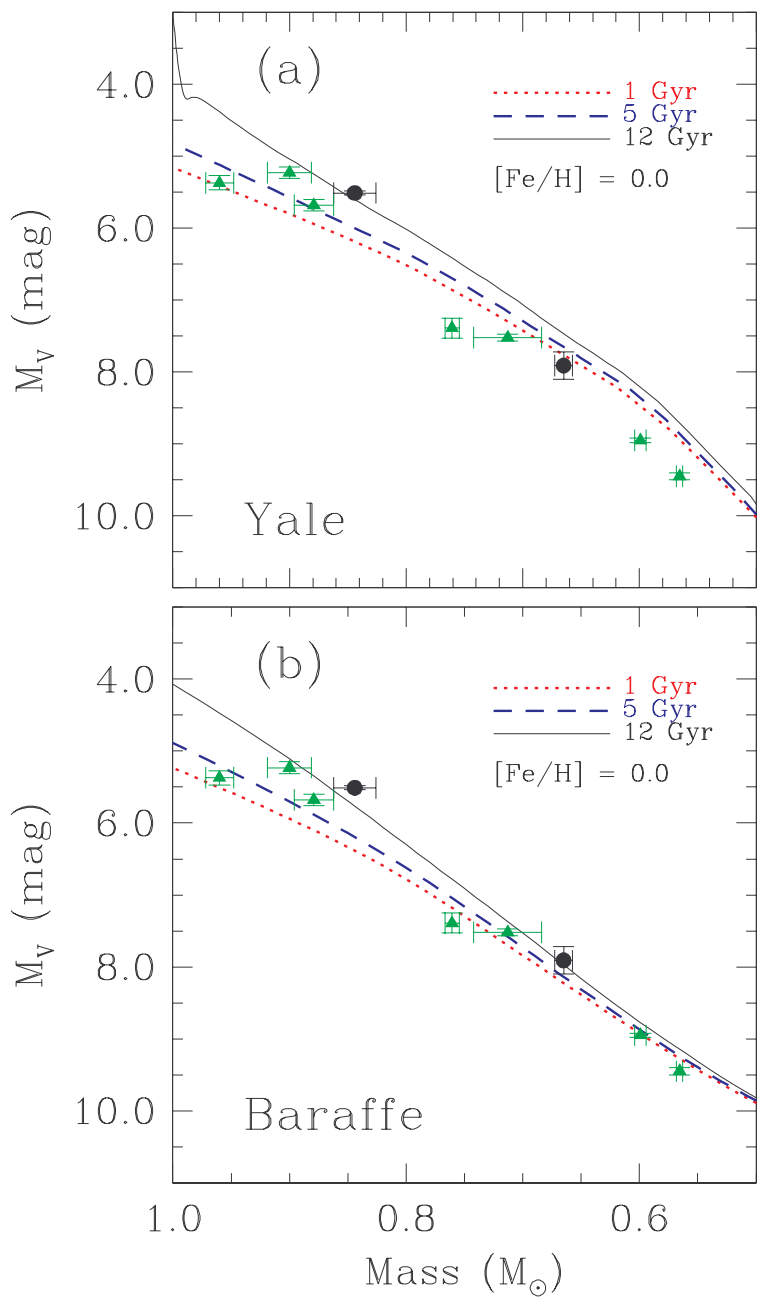}
 \figcaption[Torres.fig06.ps]{The mass-luminosity diagram in the
visual band. (a) Isochrones from the Yale-Yonsei models by Yi et al.\
(2001), as labeled, compared with the observations for HD~195987
(circles) and other binary components in this mass regime (triangles).
Infrared magnitudes have been converted to the CIT system. (b) Same as
(a), but for the models by \citet{Baraffe1998}, already on the CIT
system. The agreement is much improved.\label{fig:mvplane}}
 \end{figure}

In order to investigate the discrepancy in the absolute visual
magnitude for the secondary, we focus in Figure~\ref{fig:mvplane} on
the mass-$M_V$ plane and add observations for other binary components
that have accurately determined masses and luminosities: FL~Lyr~B,
HS~Aur~A, and HS~Aur~B \citep{Andersen1991}; Gliese~570~B and
Gliese~702~B \citep{Delfosse2000}; and V818~Tau~B and YY~Gem~AB
\citep{Torres2002}. Figure~\ref{fig:mvplane}a shows that it is not
only the secondary of HD~195987 (filled circles) that appears
underluminous compared to the \citet{Yi2001} isochrones, but four
other stars with similar masses suggest the same trend (triangles). In
Figure~\ref{fig:mvplane}b we show the same observations compared to
the Lyon models by \citet{Baraffe1998}, for the same ages and
heavy-element composition as in Figure~\ref{fig:mvplane}a. The
agreement for the secondary is now much better, and is presumably due
to the use by \citet{Baraffe1998} of sophisticated model atmospheres
as boundary conditions to the interior equations, whereas the
\citet{Yi2001} models use a gray approximation.  The latter has been
shown to be inadequate for low mass stars, where molecular opacity
becomes important \citep{Chabrier1997}. Figure~\ref{fig:mvplane}
suggests that the discrepancy begins at a mass intermediate between
that of the primary and the secondary in HD~195987.  One other
consequence of the use of the gray approximation is that the effective
temperature is typically overestimated in those models, just as hinted
by Figure~\ref{fig:yale}. 

\placefigure{fig:mvplane}

\begin{figure}
\epsscale{1.15}
\plotone{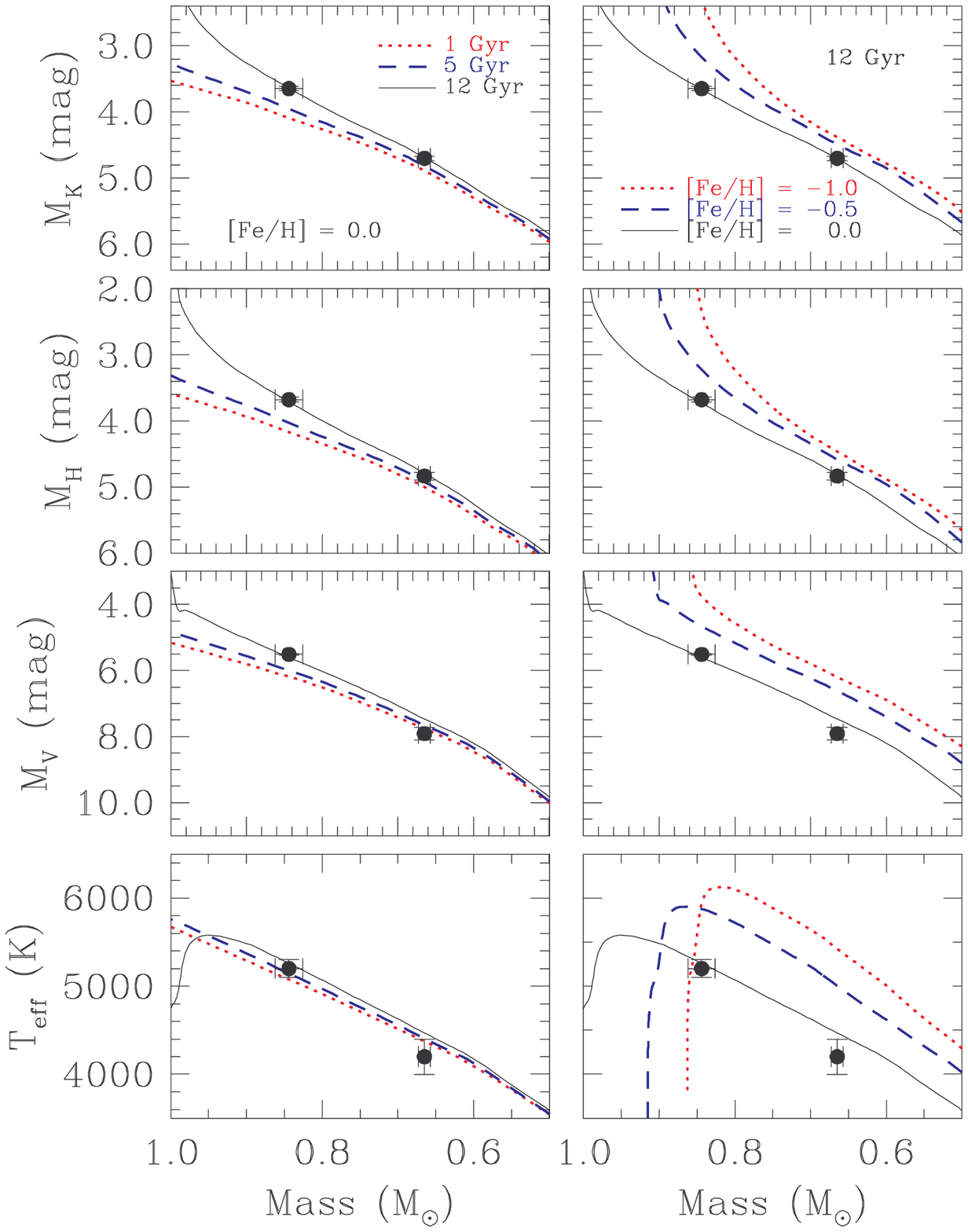}
 \figcaption[Torres.fig07.ps]{Comparison with the models by
\citet{Baraffe1998} (left) and \citet{Siess1997} (right), showing the
effect of age for solar composition, as in Figure~\ref{fig:yale}. Also
illustrated is the effect of a difference in the mixing-length
parameter $\alpha_{\rm ML}$ for the \citet{Baraffe1998} 12-Gyr
isochrones (solid lines for $\alpha_{\rm ML} = 1.0$ and dot-dashed
lines for $\alpha_{\rm ML} = 1.9$), and of a change in the
overshooting parameter $\alpha_{\rm ov}$ for the \citet{Siess1997}
12-Gyr isochrones (solid lines for $\alpha_{\rm ov} = 0.0$ and
dot-dashed lines for $\alpha_{\rm ov} = 0.2 H_{\rm
p}$).\label{fig:baraffesiess}}
 \end{figure}

In Figure~\ref{fig:baraffesiess} we illustrate the agreement between
the observations for HD~195987 and the predictions from the models by
\citet{Baraffe1998} and also by \citet{Siess1997} \citep[see
also][]{Siess2000}. The latter isochrones also use a non-gray
approximation (from fits to a different set of model atmospheres than
those used by the Lyon group), and should therefore be fairly
realistic as well.  Once again the \citet{Baraffe1998} models
(left-hand side) show reasonably good agreement in the mass-luminosity
planes for solar composition and an old age (12~Gyr), but do not quite
reproduce the estimated effective temperature of the primary. The
\citet{Siess1997} models are similar, although the agreement in $M_V$
is not as good and the fit to the primary temperature is somewhat
better. 

\placefigure{fig:baraffesiess}

In addition to the boundary conditions, another difference between the
\citet{Baraffe1998} models and the \citet{Yi2001} models is the extent
of mixing allowed, as described in the standard convection
prescription by the mixing-length parameter $\alpha_{\rm ML}$.  The
models by \citet{Yi2001} adopt a value of $\alpha_{\rm ML}=1.74$ that
best fits the observed properties of the Sun. A much lower value of
$\alpha_{\rm ML}=1.0$ is used by \citet{Baraffe1998}, whereas the best
fit to the Sun in those models requires $\alpha_{\rm ML}=1.9$.  This
has significant consequences for the temperature profile and other
properties such as the radius.  In particular, less mixing leads to a
lower effective temperature. The effect is illustrated in the diagrams
on the left side of Figure~\ref{fig:baraffesiess}, where the
dot-dashed lines gives the predictions from the \citet{Baraffe1998}
models for [Fe/H]$= 0.0$ and an age of 12~Gyr if the solar value of
$\alpha_{\rm ML}$ is used.  Changes in $M_K$, $M_H$, and $M_V$ compared
to the isochrones for the same age and metallicity but with
$\alpha_{\rm ML}=1.0$ (solid lines) are relatively small and in fact
tend to improve the agreement in $M_V$, while the temperatures
predicted with $\alpha_{\rm ML}=1.9$ are several hundred degrees
hotter than with $\alpha_{\rm ML}=1.0$ and show better agreement for
the primary (the slope is also a better match to the observations).
Presumably an intermediate value for the mixing-length parameter used
in the \citet{Baraffe1998} models would provide an optimal fit to
HD~195987 for the composition and age indicated. 

Other recent models give similar fits to the observations for
HD~195987.  For example, the recent series of isochrones from the
Padova group \citep{Girardi2000} give fits that are also good for a
metallicity near solar, and a slightly older age of 14~Gyr. 

A shortcoming of the comparisons above is that a number of additional
parameters in the isochrones are fixed, and they are somewhat
different for each series of models. One of such parameters is the
helium abundance, $Y$. The calculations by \citet{Yi2001} use $Y =
0.266$ for solar metallicity (Figure~\ref{fig:yale}), while
\citet{Baraffe1998} adopt $Y = 0.275$ for their models with
$\alpha_{\rm ML}=1.0$, but use a higher value of $Y = 0.282$ for the
more realistic models that fit the Sun with $\alpha_{\rm ML}=1.9$
(Figure~\ref{fig:baraffesiess}). \citet{Siess1997} adopted $Y =
0.277$, and \citet{Girardi2000} used $Y = 0.273$. While the
differences in these adopted helium abundances for solar metallicity
are not large, they do affect the comparisons to some degree. An
increase in $Y$ will shift the isochrones upwards in the
mass-luminosity diagrams (opposite effect as an increase in
metallicity, $Z$), and will yield slightly higher effective
temperatures. Changes in $Y$ for other metallicities are governed by
the enrichment law adopted in each series of models. 

The effect of the treatment of convective overshooting is illustrated
on the right-hand side of Figure~\ref{fig:baraffesiess} for the
\citet{Siess1997} models. The dot-dashed line corresponds to the
calculations for an age of 12~Gyr and an overshooting parameter of
$\alpha_{\rm ov} = 0.2 H_{\rm p}$ (where $H_{\rm p}$ is the pressure
scale height), while the other isochrones assume no overshooting. The
change compared to the 12~Gyr isochrone with no overshooting (solid
line) is hardly noticeable. 

\subsection{The iron abundance of HD~195987}
\label{sec:iron}

The model comparisons above seem to point toward a metallicity for the
system near solar that is at odds with the observations, along with a
fairly old age ($\sim$10--12~Gyr), perhaps a somewhat unusual
combination.  Because of the possible implications of this
disagreement for our confidence in the models, in this section we
examine each metallicity determination more closely in an effort to
understand these discrepancies, and we attempt to quantify possible
systematic errors. 

The presence of the secondary, even though it is faint, affects the
total light of the system at some level introducing subtle biases in
the photometric estimates of the metallicity of HD~195987, which are
derived from the combined colors assuming that they correspond to a
single star. The photometric estimates of the temperature based on the
same assumption are also affected, and this may propagate through and
affect some of the spectroscopic abundance determinations as well. 

In order to quantify these effects we have simulated binary systems by
combining the single-star photometry of a primary and a secondary,
each with normal colors, based on the same tabulations used in
\S\ref{sec:physics}. We computed various photometric indices from the
combined light, and used them to estimate the metallicity of HD~195987
following the same procedures employed by \citet{Marsakov1988} (who
relied on the ultraviolet excess in the Johnson system) and
\citet{Wyse1995} (based on Str\"omgren photometry).  We then compared
these results with those obtained for the primary alone, for a range
of magnitude differences $\Delta V$ between the primary and secondary.
In all cases we selected the primary so as to reproduce the actual
observed colors of HD~195987 in each photometric system at $\Delta V =
2.4$~mag. 

\begin{figure}
\epsscale{0.9}
\plotone{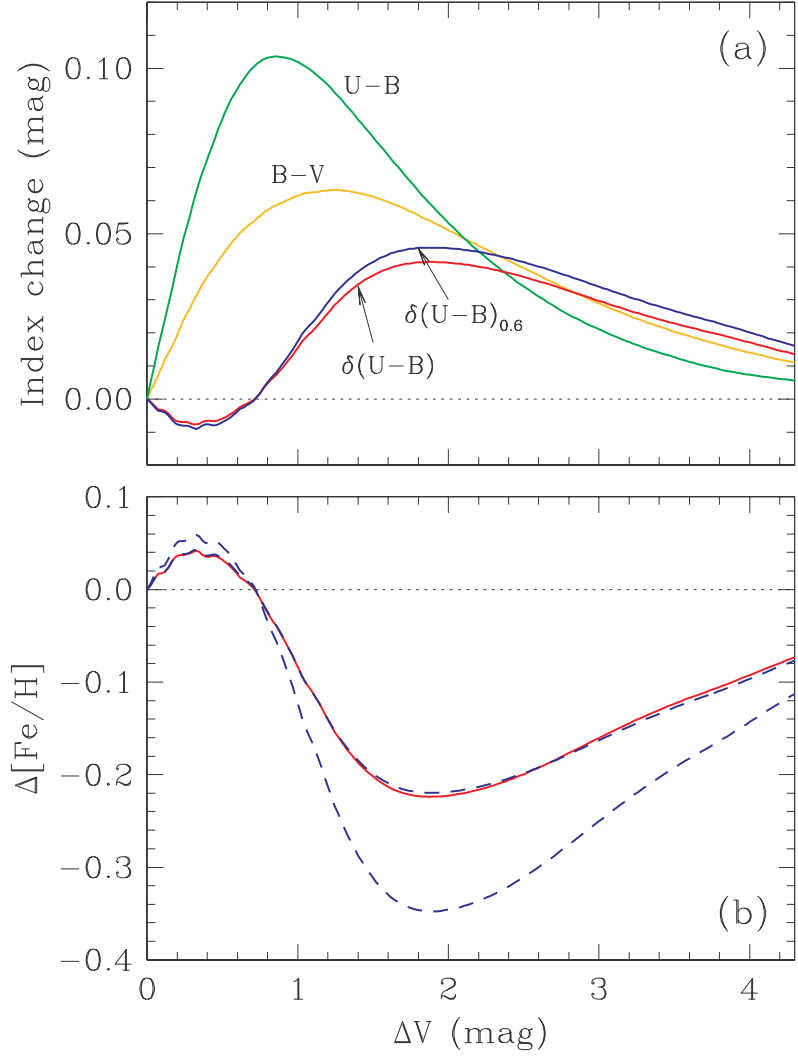}
 \figcaption[Torres.fig08.ps]{Effect of the light from the secondary
in a binary, as a function of the magnitude difference in the visual
band. (a) Changes in the color indices in the Johnson system,
including the ultraviolet excess with and without correction for the
guillotine \citep{Sandage1969}; (b) Impact on the metallicity
determination when using the ultraviolet excess $\delta(U\!-\!B)$ and
the calibration by \citet{Marsakov1988} (solid line), or the
normalized excess $\delta(U\!-\!B)_{0.6}$ \citep{Sandage1969} and
calibrations by \citet{Carney1979} (dashed lines).\label{fig:johnson}}
 \end{figure}

Figure~\ref{fig:johnson}a shows how the presence of the secondary
affects the $U\!-\!B$ and $B\!-\!V$ colors, as well the ultraviolet
excess $\delta(U\!-\!B)$ used by \citet{Marsakov1988}, as a function
of the magnitude difference. The normalized ultraviolet excess
$\delta(U\!-\!B)_{0.6}$ \citep[corrected for the guillotine;
see][]{Sandage1969}, also frequently used for metallicity estimates,
has a similar behavior. Figure~\ref{fig:johnson}b illustrates the
effect on the derived metallicity using the calibration by
\citet{Marsakov1988}, and also two calibrations by \citet{Carney1979}
based on $\delta(U\!-\!B)_{0.6}$.  The [Fe/H] estimates are biased
toward \emph{lower} values\footnote{The sign of these changes may seem
counterintuitive. The secondary star tends to redden both the
$U\!-\!B$ and the $B\!-\!V$ indices, which ought to make the combined
light of the system appear more metal-rich. As it turns out, the slope
of this ``reddening" vector in the $U\!-\!B$ vs.\ $B\!-\!V$ plane is
smaller than that of the blanketing vector, so that the net effect is
a bias toward \emph{larger} ultraviolet excesses, and therefore lower
metallicity estimates. We find, however, that in general this depends
both on the color of the primary and on the magnitude difference
$\Delta V$. For bluer primaries than that of HD~195987 the sign of the
change is reversed in certain ranges of $\Delta V$.}, and the maximum
effect is for a secondary about 2 magnitudes fainter than the primary.
If the original determination by \citet{Marsakov1988} ([Fe/H]$=-0.34$)
is corrected for this effect ($\sim$0.2~dex at $\Delta V=2.4$~mag),
the result is much closer to the solar abundance. 

\placefigure{fig:johnson}

The normalized ultraviolet excess observed for HD~195987 is
$\delta(U\!-\!B)_{0.6} = 0.061$. This leads to [Fe/H]$=-0.13$ or
[Fe/H]$=-0.18$ using the linear or quadratic calibration formulae by
\citet{Carney1979}. From Figure~\ref{fig:johnson}b it is seen that in
this case the corrected values would be [Fe/H]$\sim +0.1$. 

\begin{figure}
\epsscale{0.9}
\plotone{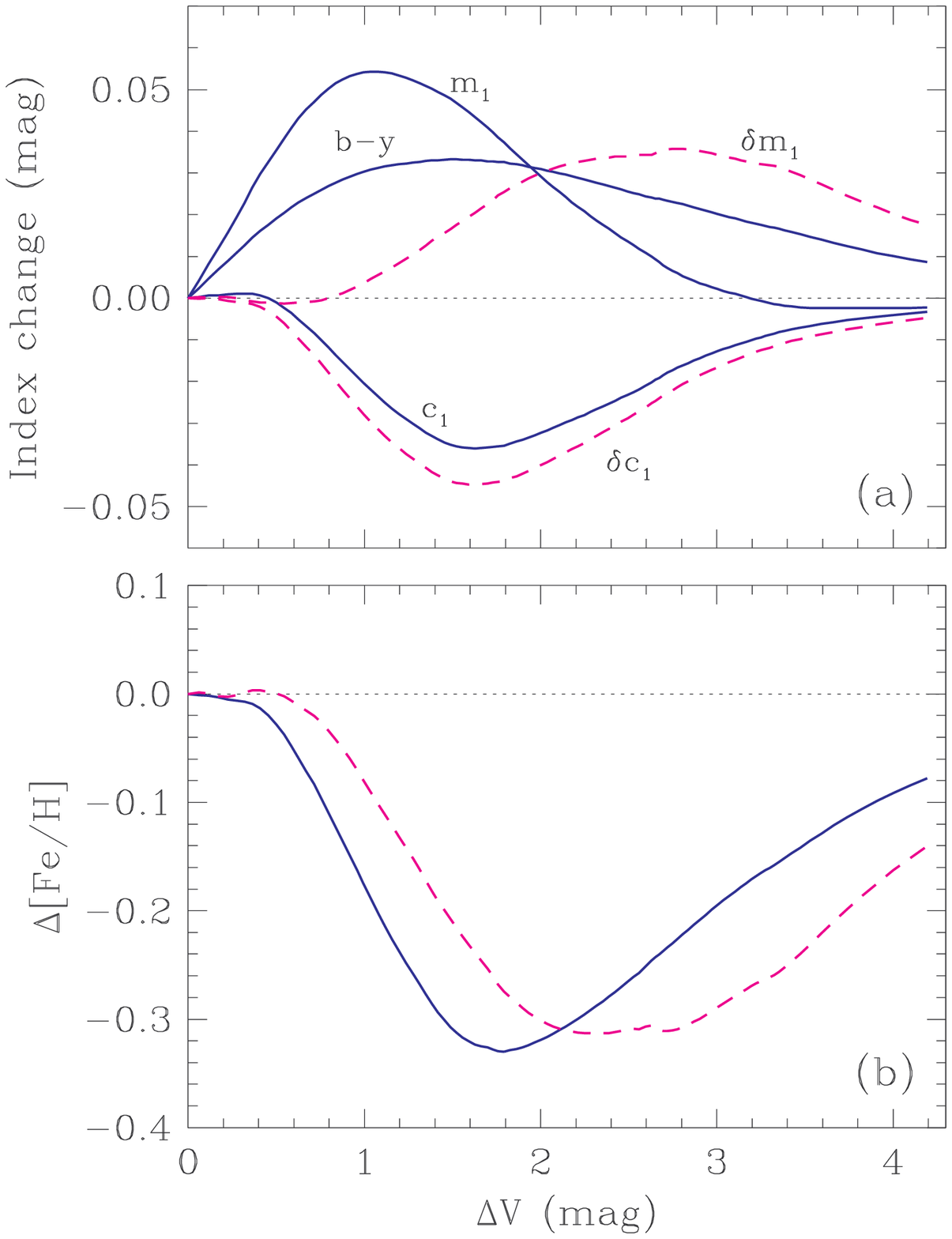}
 \figcaption[Torres.fig09.ps]{Effect of the light from the secondary
in a binary, as a function of the magnitude difference in the visual
band. (a) Changes in the color indices in the Str\"omgren system. The
indices represented with solid lines are used in the metallicity
calibration by \citet{Schuster1989} adopted by \citet{Wyse1995}, and
the ones shown as dashed lines are used in the calibration by
\citet{Olsen1984}; (b) Impact on the metallicity determination when
using the calibration by \citet{Schuster1989} (solid line) and by
\citet{Olsen1984} (dashed).\label{fig:stromgren}}
 \end{figure}

Similar effects occur in the Str\"omgren system. The changes in the
photometric indices are shown in Figure~\ref{fig:stromgren}a.  The
estimate of [Fe/H]$ = -0.31$ by \citet{Wyse1995} is based on the
quantities \{$b\!-\!y$, $m_1$, $c_1$\} and the calibration by
\citet{Schuster1989}.  A different calibration by \citet{Olsen1984}
involving \{$\delta m_1$, $\delta c_1$\} yields a nearly identical
value of [Fe/H]$=-0.32$.  As seen in Figure~\ref{fig:stromgren}b the
bias toward lower metallicities due to the secondary happens to reach
a maximum quite near the $\Delta V$ of our system. If accounted for,
the resulting metal abundance from the Str\"omgren indices would be
essentially solar. 

\placefigure{fig:stromgren}

Though it may seem that these corrections bring the photometric
estimates of the metallicity of HD~195987 in line with the indications
from the model comparisons described earlier, the photometric
determinations of [Fe/H] are as a rule less trustworthy than those
based on high-resolution spectroscopy \citep[see,
e.g.,][]{Gehren1988}.

The spectroscopic determinations in turn are not, however, completely
insensitive to the presence of the secondary either. Two different
effects must be considered.  On the one hand the continuum from the
secondary typically tends to fill in the spectral lines of the
primary, which then appear weaker as if the star were more metal-poor.
On the other hand the combined-light photometry is reddened.
Therefore, the temperature estimates that are used to begin the
spectroscopic analysis are biased toward lower values. 

The metallicity determination by \citet{Carney1994}, [m/H]$= -0.60$,
which supersedes that by \citet{Laird1988}, is based on a subset of
the same spectra that we have used in this paper, which are of high
resolution but low S/N.  To derive the metal abundance they used a
$\chi^2$ technique to compare the observations against a grid of
synthetic spectra for a range of metallicities, adopting a fixed
temperature determined from photometric indices. As they discuss, the
[m/H] values derived for double-lined binaries with this method can be
quite different depending on whether the spectra that are used have
the lines of the two components well separated or exactly aligned.  In
the first case the continuum from the secondary tends to fill in the
lines of the primary, as mentioned earlier, and the metallicity is
biased toward lower values. In the second case, when the velocities of
the stars are similar (close to the center-of-mass velocity $\gamma$
of the binary), the effect is the opposite because the average line
strength is slightly increased due to the contribution of the cooler
secondary.  \citet{Carney1994} used only the few spectra near the
$\gamma$ velocity available to them, because then the bias toward
higher metallicities is balanced to some degree by the bias toward
lower metallicities resulting from the lower temperature inferred from
the combined light.  The latter is the result of the strong
correlation between metallicity and temperature, which is essentially
the same as that mentioned in \S\ref{subsec:spectroscopy}. 

\begin{figure}
\epsscale{0.9}
\plotone{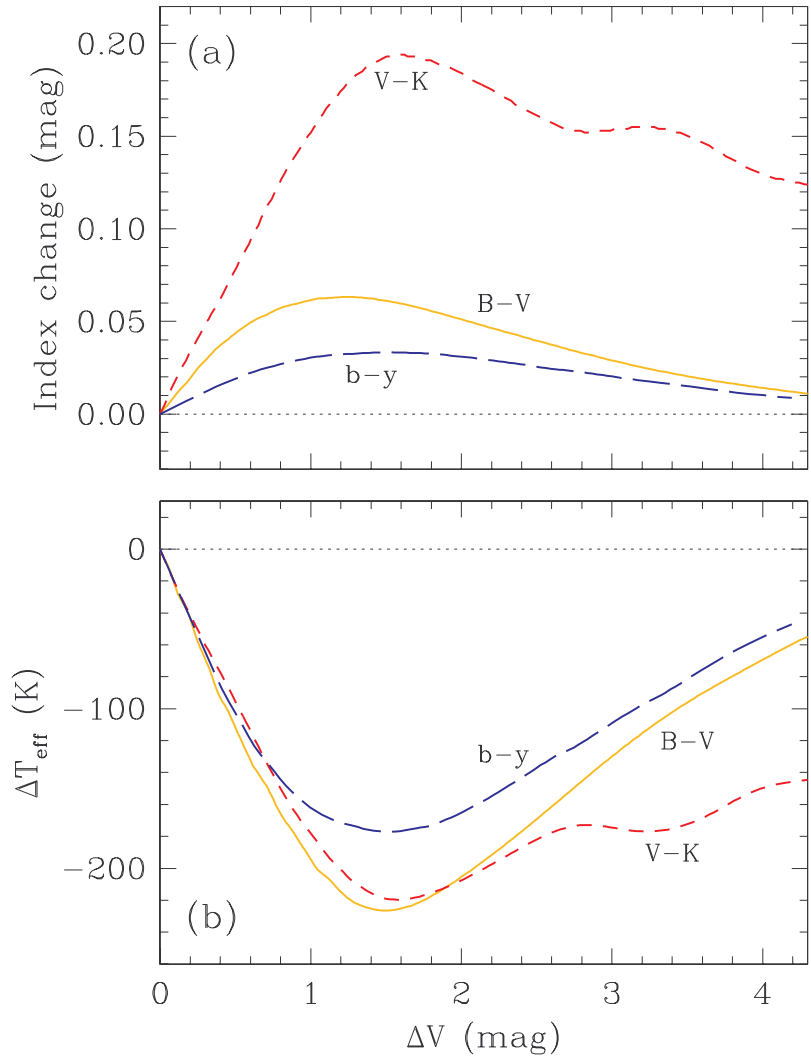}
 \figcaption[Torres.fig10.ps]{Effect of the light from the secondary
as a function of the magnitude difference in the visual band. (a)
Changes in the color indices, as labeled; (b) Bias in the temperature
determinations using $B\!-\!V$, $V\!-\!K$, and $b\!-\!y$, and the
calibrations by \citet{Carney1994}.\label{fig:teff}}
 \end{figure}

In Figure~\ref{fig:teff}a we illustrate the magnitude of the effect of
the secondary on the broadband and intermediate-band color indices
$B\!-\!V$, $b\!-\!y$, and also $V\!-\!K$, from simulations analogous
to those performed earlier. The contamination is much more noticeable
in $V\!-\!K$, of course, because the secondary becomes comparatively
brighter in $K$. Changes in the effective temperature inferred from
the combined light compared to those for the primary alone are shown
in Figure~\ref{fig:teff}b, based on the same calibrations used by
\citet{Carney1994}. 

\placefigure{fig:teff}

Simulations performed by \citet{Carney1994} to estimate the residual
bias on the metallicity for double-lined systems indicate that in
general the two effects do not quite cancel each other out. We
estimate from their experiments that their metal abundance [m/H] for
HD~195987 may still be too low by 0.15--0.20 dex, accounting for the
fact that this system is a bit more extreme than the simulated
binaries they considered.  Note also that this metallicity estimate is
labeled [m/H] instead of [Fe/H] because it includes all metals with
lines present in the spectral window. The distinction could in
principle be an important one due to the presence of the strong
\ion{Mg}{1}~b lines in the CfA spectra, and the typically enhanced
abundance of this element relative to iron in metal-poor stars (see
next section).  Tests described by \citet{Carneyetal1987} indicate
that the correction required for this effect is not likely to be more
than about $-0.05$~dex for this system, especially given that it may
be partially masked by other corrections they applied to the
abundances based on comparisons with metallicity standards that may
also have enhanced magnesium (J.\ Laird \& B.\ Carney 2002, priv.\
comm.). The contribution of the two effects described above leads to
an adjusted \citet{Carney1994} estimate of [Fe/H] between $-0.50$ and
$-0.45$. 

The metallicity determination by \citet{Beers1999}, [Fe/H]$= -0.52$,
is based on lower-resolution spectra (1--2~\AA) and relies on both the
strength of a \ion{Ca}{2}~K-line index and on the height of the peak
of the Fourier autocorrelation function of the spectrum, which is a
line strength indicator.  Both indices are calibrated against the
observed $B\!-\!V$ so as to remove the temperature dependence.  The
unrecognized presence of the secondary is unlikely to weaken the
spectral lines of the primary very much in this case since the
observations cover only the blue portion of the spectrum, where the
secondary is fainter. However the reddening in the $B\!-\!V$ color,
estimated to be $\sim$0.04~mag from Figure~\ref{fig:johnson}a, may
have some effect on [Fe/H]. It is difficult to quantify its effect due
to the complexity of the technique used by \citet{Beers1999}, but on
the basis of their Figure~6 and Figure~9 the bias will be toward
lower abundances for both of their metallicity indices, and a
correction of $+0.1$~dex or more may be appropriate. With this
adjustment the value for HD~195987 would be [Fe/H]$\approx -0.45$ to
$-0.40$. 

Finally, the detailed high-resolution ($\lambda/\Delta\lambda \sim
50,\!000$) abundance analysis by \citet{Fulbright2000} gives
[Fe/H]$=-0.66$, but also does not account for the presence of the
secondary star. The measured equivalent widths and consequently the
metal abundances for iron and other elements are thus expected to be
slightly underestimated due to light of the secondary filling in the
lines of the primary, given that their spectrum was obtained at an
orbital phase when the lines were not exactly aligned. The magnitude
of the effect may depend on the strength of the lines used in the
analysis, and could be as large as 0.1~dex. In addition,
\citet{Fulbright2000} adopted an initial estimate of the effective
temperature based on the $V\!-\!K$ index, which according to
Figure~\ref{fig:teff}b leads to an underestimate of $\Delta T_{\rm
eff}\sim$180~K due to the infrared excess. Although their use of iron
lines of different strength (to constrain the microturbulence),
different excitation potentials (to constrain $T_{\rm eff}$), and
different ionization stages (to contrain $\log g$) should tend to
reduce the sensitivity of the results to this problem, the parameter
sensitivity experiments in Table~8 by \citet{Fulbright2000} indicate
that such an error in $T_{\rm eff}$ can still produce changes in the
resulting abundance of 0.07--0.16~dex (depending on whether other
parameters are held fixed or allowed to vary to compensate for $\Delta
T_{\rm eff}$).  The changes will be in the direction of lower
abundances for lower temperatures.  The combined effect of these
corrections results in a metallicity of [Fe/H]$\approx -0.50$ to
$-0.40$. 

It may seem that many of these adjustments to account for the
secondary are rather small and perhaps even within the uncertainties
of some of the determinations. Indeed they are, but they are also
\emph{systematic} in nature and \emph{known to be present}.  And
because we are trying to understand a systematic difference between
the estimates of [Fe/H] and indications from all the models, these
biases cannot be ignored, particularly since they often go in the same
direction.  To summarize the state of the (corrected) metallicity
estimates for HD~195987, it appears that both of the photometric
determinations result in abundances very near solar, while the three
spectroscopic estimates, which should be more reliable, yield [Fe/H]
values between $-0.50$ and $-0.40$. These are still decidedly lower
than the solar value, and so the apparent disagreement with the
evolutionary models persists if we give preference to the
spectroscopic abundances. 

\subsection{Detailed abundances for other elements; further model
comparisons}
\label{sec:alpha}

The detailed chemical analysis by \citet{Fulbright2000} has revealed
certain patterns in HD~195987 that are quite typical in metal-poor
stars, and that turn out to be key to understanding the comparison
with theoretical isochrones. In particular, it is well known that the
abundance of the so-called $\alpha$-elements (O, Ne, Mg, Si, S, Ar,
Ca, and Ti) in metal-poor stars is usually enhanced relative to iron
when compared to the same ratios in Sun
\citep{Conti1967,Greenstein1970}.  The enhancement, [$\alpha$/Fe], is
actually observed to depend on the metallicity as a result of
enrichment from Type II and Type Ia supernovae on different timescales
throughout the history of the Galaxy \citep[see,
e.g.,][]{Wheeler1989}. The average enhancements for the different
$\alpha$-elements increase from the solar values at [Fe/H]$= 0.0$ to
about $+0.3$ to $+0.5$~dex at [Fe/H]$\sim -1$ or so, and then they
remain approximately constant for more metal-deficient stars.  There
is even evidence that different stellar populations in the Galaxy are
chemically distinct \citep[see,
e.g.,][]{Edvardsson1993,Mashonkina2000,Prochaska2000}.  Four of the
$\alpha$-elements in HD~195987 have been measured by
\citet{Fulbright2000}, and have been found to be enhanced.  The ratios
measured are [Mg/Fe]$=+0.44 \pm 0.07$, [Si/Fe]$=+0.41 \pm 0.07$,
[Ca/Fe]$=+0.26 \pm 0.07$, and [Ti/Fe]$=+0.31 \pm 0.09$ in the standard
logarithmic notation. 

A stellar mixture rich in these elements produces an increase in the
opacity that affects the structure of a star and changes both the
temperature and the luminosity predicted by the models. Compared to
model calculations with solar ratios for all elements,
$\alpha$-enhancement leads to cooler temperatures and lower
luminosities. Therefore, these anomalies must be accounted for if a
proper interpretation is to be made of the observations.  A number of
authors have noted, however, that these effects can be mimicked in
models that assume solar ratios by simply increasing the overall
metallicity used to compute the tracks \citep[see,
e.g.,][]{Chieffi1991,Chaboyer1992,Baraffe1997,VandenBerg2000}.
Prescriptions for how to do this as a function of the
$\alpha$-enhancement have been presented by \citet{Salaris1993} and
\citet{VandenBerg2000}, among others.  In the absence of a detailed
measurement of the abundance of the $\alpha$-elements for a particular
object, average values of [$\alpha$/Fe] have often been adopted in
different ranges of [Fe/H].  This carries some risk, however, given
that there is considerable scatter in the measured enhancements for
different stars at any given [Fe/H] value
\citep{Edvardsson1993,Carney1996,Prochaska2000}. 

In the case of HD~195987 we have a direct measurement of [$\alpha$/Fe]
if we assume that all the $\alpha$-elements follow the trend of Mg,
Si, Ca, and Ti, which were the ones actually measured.  This is
generally observed to be true to first order in metal-poor stars.  The
average overabundance for the four measured elements is [$\alpha$/Fe]$
= +0.36 \pm 0.12$\footnote{Because we are dealing in this case with
\emph{ratios} of element abundances, the effect that the secondary
component might have on those determinations is not expected to be as
significant as in the case of [Fe/H]. Table~8 by \citet{Fulbright2000}
indicates that the bias due to temperature errors averaged over Mg,
Si, Ca, and Ti (which have different signs for the correlation with
$T_{\rm eff}$) is 0.03~dex at most, with the sign depending on whether
only the temperature is changed or whether other fitted parameters are
allowed to vary simultaneously to compensate. Similarly, the continuum
from the secondary should have a minimal effect on the abundance
ratios.}. Following \citet{VandenBerg2000}, the overall metallicity
adjustment required for this level of [$\alpha$/Fe] is $+0.27$~dex.
Given the estimates of [Fe/H] in the range $-0.50$ to $-0.40$
(\S\ref{sec:iron}), one would then expect that model isochrones
computed for a metallicity between $-0.23$ and $-0.13$ should provide
reasonably good fits to the observations.  This is in fact quite close
to what we find, as shown by the comparisons in \S\ref{sec:modelcomp},
which suggested an overall composition between solar and $-0.15$. 

\begin{figure}
\epsscale{0.9}
\plotone{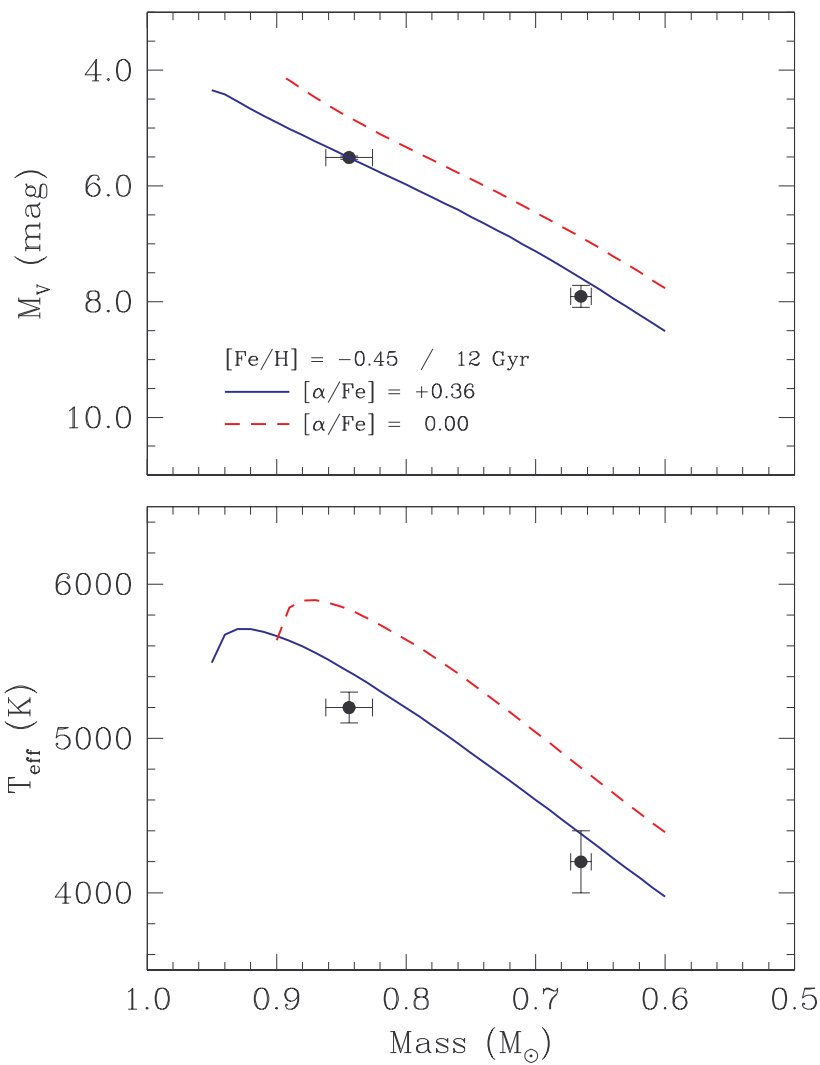}
 \figcaption[Torres.fig11.ps]{Isochrone fits based on the models by
\citet{Bergbusch2001} that include $\alpha$-element enhancement. a)
The fit shown is to the primary value of $M_V$ only (see text), for an
age of 12~Gyr and a fixed value of [$\alpha$/Fe]$ = +0.36$. The iron
abundance from the fit is [Fe/H]$ = -0.45$. A similarly good fit is
obtained for an age of 10~Gyr and [Fe/H]$ = -0.52$. Also represented
for comparison is a model for the same [Fe/H] but with [$\alpha$/Fe]$
= 0.0$ (dashed line). b) Effective temperature predictions from the
same models.  \label{fig:vanden}}
 \end{figure}

Though adjusting the metallicity of models with scaled-solar mixtures
may mimic those with $\alpha$-enhancements to a good approximation,
the match is not perfect, as shown by \citet{VandenBerg2000}.
Isochrones incorporating $\alpha$-enrichment for a range of values of
[$\alpha$/Fe] between 0.0 to $+0.6$ have recently been published by
\citet{Bergbusch2001}. We compare them with the observations for
HD~195987 in Figure~\ref{fig:vanden}, in the $M_V$ vs.\ mass and
$T_{\rm eff}$ vs.\ mass diagrams ($H$- and $K$-magnitude predictions
are not available in this series of calculations). Because these
models use a gray approximation for the boundary conditions between
the photosphere and the interior, we do not expect them to reproduce
the brightness of the secondary or the temperature of either star very
well (see \S\ref{sec:modelcomp}). Thus we must rely only on the
absolute visual magnitude of the primary component for this test. 

\placefigure{fig:vanden}

For a fixed value of [$\alpha$/Fe]$= +0.36$ the iron abundance that
provides the best match is [Fe/H]$ = -0.45$ for an age of 12~Gyr
(Figure~\ref{fig:vanden}), and [Fe/H]$ = -0.52$ for 10~Gyr. These
values are in excellent agreement with the measured abundances by
\citet{Carney1994}, \citet{Beers1999}, and \citet{Fulbright2000},
after correcting for the biases due to the secondary as described
before. The effect of the $\alpha$-enhancement is illustrated in the
figure by the shift between the solid line ([$\alpha$/Fe]$= +0.36$)
and the dashed line ([$\alpha$/Fe]$ = 0.0$) for the same metallicity.
The uncertainty in the adopted value of [$\alpha$/Fe] (0.12~dex)
translates into the same uncertainty for the fitted value of [Fe/H].
The discrepancy for the $M_V$ of the secondary and for the
temperatures is in the same direction as shown earlier, i.e., the
models are overluminous for a star with the mass of the secondary, and
both temperatures are overestimated. 

In addition to the $\alpha$-elements, abundance ratios relative to
iron were measured by \citet{Fulbright2000} for the light elements Na
and Al, the iron-peak elements V, Cr, and Ni, and the heavy elements
Ba, Y, and Eu. A comparison with similar measurements for other stars
reported by \citet{Prochaska2000} reveals that the pattern of
enhancements seen in HD~195987 is very similar to that exhibited by
thick disk stars, and is clearly distinct from that of the thin disk
stars \citep[see also][]{Fuhrmann1998}. On that basis one might
conclude that HD~195987 belongs to the thick disk population of our
Galaxy.  This would typically imply a rather old age for the binary,
in qualitative agreement with the indications from the models in
\S\ref{sec:modelcomp}.  The kinematics of the system computed from the
proper motion, the systemic radial velocity, and our orbital parallax
(Galactic velocity components $U = 18~\kms$, $V = 13~\kms$, $W =
48~\kms$, relative to the Local Standard of Rest) lend some support to
this idea: the $W$ velocity is quite consistent with thick disk
membership \citep[see][]{Carney1989}, although the $V$ component is
not as extreme. 

The abundances of Eu and Ba are of particular astrophysical interest.
Europium in the solar system is produced mostly by rapid neutron
capture (``\emph{r}-process"), for which the principal formation site
is believed to be Type II supernovae. It is one of the few
\emph{r}-process elements with relatively unblended atomic lines
present in the visible part of the spectrum, which makes it very
useful as an indicator for the \emph{r}-process history of stellar
material.  Barium, on the other hand, is predominantly a slow neutron
capture (``\emph{s}-process") element that is thought to be produced
mainly during the thermal pulses of asymptotic giant branch stars.
The abundance ratio between Eu and Ba is therefore particularly
sensitive to whether nucleosynthesis of heavy elements occured by the
\emph{s-} or \emph{r}-process, and as such it is a useful diagnostic
for studying the chemical evolution of the Galaxy.  [Eu/Ba] is
observed to increase with metal deficiency, and has also been shown to
correlate quite well with age \citep{Woolf1995}. For HD~195987
\citet{Fulbright2000} measured [Eu/Fe]$ = +0.29 \pm 0.10$ and [Ba/Fe]$
= -0.10 \pm 0.12$, so that [Eu/Ba]$ = +0.39 \pm 0.14$. Once again this
is quite typical of the values for thick disk stars
\citep[see][]{Mashonkina2000}.  Furthermore, according to the
correlation reported in Figure~14 by \citet{Woolf1995} this ratio
suggests an age for HD~195987 that is at least 10--12~Gyr, in agreement
with indications from the stellar evolution models used here and also
consistent with the old age of the thick disk \citep{Carney1989}. 

\subsection{The overall agreement with the models}
\label{sec:overall}

The comparisons in \S\ref{sec:modelcomp} and \S\ref{sec:alpha}
indicate that current stellar evolution theory can successfully
predict all of the observed characteristics of HD~195987 to high
precision for a chemical composition matching what is measured.
Unfortunately, however, none of the published models is capable of
giving a good fit to all properties of both stars simultaneously. The
various isochrones used in this paper do at least agree in predicting
an old age for the system, which is also consistent with the pattern
of enhancement of the heavy elements, as described above. 

The low luminosity of the secondary star in the visual band lends
strong support to models that use sophisticated boundary conditions to
the interior equations, as in \citet{Baraffe1998}, particularly for
the lower main sequence.  This also appears to be required in order to
reproduce the effective temperatures of these stars, but not with a
mixing-length parameter as low as in those models. A value of
$\alpha_{\rm ML}$ much closer to the one for the Sun \citep[as used in
most models \emph{except} those by][]{Baraffe1998} seems indicated. 

The enhanced levels of the $\alpha$-elements in HD~195987 and in other
metal-poor stars cannot be ignored in the models, although their
effect can be approximated in standard calculations by increasing the
overall metallicity. Note, however, that this approach defeats the
true purpose of the model comparisons, which is to test whether theory
can reproduce all observed properties at the measured value of [Fe/H].
In the absence of a detailed chemical analysis for metal-poor stars,
the constraint on models is considerably weakened because a
satisfactory fit can usually be found by leaving the metallicity as a
free parameter.  The models by \citet{Bergbusch2001} incorporate the
effect of $\alpha$-enhancements, but regretably not the more refined
boundary conditions used by \citet{Baraffe1998}, which probably
explains their failure to reproduce the $M_V$ of the secondary and the
effective temperatures of both stars. 

The example of HD~195987 emphasizes the importance of detailed
spectroscopic analyses as a crucial ingredient for proper comparisons
of metal-poor stars with stellar evolution theory. Not only is an
accurate determination of [Fe/H] needed, but also abundances for other
elements, particularly the $\alpha$-elements that contribute the most
to the opacities of these objects.  To our knowledge the binary
studied in this paper represents the most metal-poor object with such
an analysis that has absolute masses determined to 2\% or better.
HD~195987 is perhaps a favorable case for abundance determinations,
however, because the secondary is faint enough that it does not
interfere significantly with that analysis.  This is seldom the case
in double-lined eclipsing binaries that have accurately determined
physical properties. 

While the tests described in previous sections appear to validate
stellar evolution theory, the constraint is actually not as strong as
it could be because the absolute radii for the components are not
directly measured (but see below).  The stellar radius is a very
sensitive measure of evolution (age), and the fact that it can be
derived by purely geometrical means in binaries that are eclipsing is
what makes such systems so valuable.  From accurate observations of a
number of systems that do eclipse, there is now clear evidence that
while current models are quite successful in predicting the radiative
properties of stars, they fail to reproduce the radii of stars in the
lower main-sequence
\citep[see][]{Popper1997,Lastennet1999,Clausen1999}, predicting sizes
that are too small.  The discrepancy can be as large as 10--20\%, as
demonstrated recently for YY~Gem~AB and V818~Tau~B by
\citet{Torres2002}. This may have a significant impact on ages
inferred from these models. 

Incidentally, we note that the absolute radii that we derive in
\S\ref{sec:physics} for the components of HD~195987 based on the
radiative properties of the system are also somewhat larger than
predicted for stars of this mass by all of the models considered here,
by roughly 10\%.  Though more uncertain and possibly not as reliable
as the geometric determinations that might be obtained in eclipsing
systems, our estimates are consistent with the trend described above. 

\section{Final remarks}
\label{sec:finalremarks}

New spectroscopic observations of the nearby moderately metal-poor
double-lined spectroscopic binary system HD~195987 ([Fe/H]$\sim -0.5$,
corrected for binarity), along with interferometric observations that
clearly resolve the components for the first time, have allowed us to
obtain accurate masses ($\sigma\leq 2$\%), the orbital parallax, and
component luminosities in $V$, $H$, and $K$.  Both the detailed
chemical composition (including the pattern of enhancement of the
$\alpha$-elements and heavier elements such as Ba and Eu) and the
kinematics suggest that the system is a member of the thick disk
population of our Galaxy, and an age of 10--12~Gyr or more (consistent
with that conclusion) is inferred based on isochrone fits. 

The determination of these stellar properties places useful
constraints on stellar evolution theory for the lower main sequence.
We have shown that none of the models considered here fits all the
properties for both components simultaneously, although the
discrepancies in each case can be understood in terms of the physical
assumptions (convection prescriptions, boundary conditions, detailed
chemical composition). Indications are that a model incorporating all
the proper assumptions \emph{together} (which is not done in any
available series of calculations) would allow for a good match to the
observations of HD~195987. While this may not be straightforward to do
in practice due to the complexity of the problem, it would seem to be
an obvious action item for theorists, which would enable observational
astronomers to perform more useful tests for stars under 1~M$_{\sun}$.
The ingredients identified here as making the most significant
difference in the fit for the mass regime of this binary are: (a) the
use of non-gray boundary conditions between the photosphere and the
interior, based on modern model atmospheres that incorporate molecular
opacity sources; (b) a mixing-length parameter close to that required
for the Sun; and (c) the inclusion of the effect of enhanced
abundances of the $\alpha$-elements. 

Though it seems that all the radiative properties of HD~195987 can
in principle be predicted accurately from current theory at the
observed metallicity and $\alpha$-enhancement, there is no constraint
in this particular case on the radius, a key diagnostic of evolution.
Evidence cited earlier from other studies indicates that some
adjustments are still needed in the models, which tend to
underestimate the sizes of low-mass stars. 

Further progress on the observational side in testing models for
metal-poor stars requires additional candidates suitable for accurate
determinations of their physical properties. The recent paper by
\citet{Goldberg2002} presents a list of nearly 3 dozen double-lined
spectroscopic binaries from a proper motion sample, many of which are
metal-deficient. Improvements in the sensitivity of ground-based
interferometers in the coming years should allow several of them to be
spatially resolved. Eclipsing binary candidates found in globular
clusters in a number of recent studies are also valuable, but are
fainter and the spectroscopy will be more challenging. In both cases
detailed chemical analyses will be very important for a meaningful
test of theory, as shown by the system reported here. 
	
\acknowledgements 

We are grateful to Joe Caruso, Bob Davis, and Joe Zajac for obtaining
many of the spectroscopic observations, and to Bob Davis for also
maintaining the CfA echelle database. We thank Lucas Macri for
assistance in obtaining the $H$ magnitude of HD~195987, and Bruce
Carney and John Laird for a careful reading of the manuscript and for
very helpful comments on abundance issues.  The anonymous referee is
also thanked for a number of useful suggestions.  Some of the
observations reported here were obtained with the Multiple Mirror
Telescope, a joint facility of the Smithsonian Institution and the
University of Arizona.  Interferometer data were obtained at Palomar
Observatory using the NASA Palomar Testbed Interferometer (PTI),
supported by NASA contracts to the Jet Propulsion Laboratory.  Science
operations with PTI are conducted through the efforts of the PTI
Collaboration ({\tt
http://huey.jpl.nasa.gov/palomar/ptimembers.html}), and we acknowledge
the invaluable contributions of our PTI colleagues and the PTI
professional observer Kevin Rykoski.  MP acknowledges support from the
National Science Foundation Research Experiences for Undergraduates
program at SAO. This research has made use of software produced by the
Interferometry Science Center at the California Institute of
Technology.  This research has also made use of the SIMBAD database,
operated at CDS, Strasbourg, France, of NASA's Astrophysics Data
System Abstract Service, of the Washington Double Star Catalog
maintained at the U.S.\ Naval Observatory, and of data products from
the Two Micron All Sky Survey, which is a joint project of the
University of Massachusetts and the Infrared Processing and Analysis
Center, funded by the National Aeronautics and Space Administration
and the National Science Foundation.

\clearpage

\clearpage

\clearpage

\clearpage

\clearpage

\clearpage

\clearpage

\clearpage

\clearpage

\clearpage

\clearpage

\clearpage

\begin{deluxetable}{crrrrc}
\tablenum{1}
\tablewidth{26pc}
\tabletypesize{\scriptsize}
\tablecaption{Radial velocity measurements of HD~195987 and residuals from the final fit.\label{tab:rvs}}
\tablehead{
\colhead{HJD} & \colhead{RV$_{\rm A}$} & \colhead{RV$_{\rm B}$} &
\colhead{(O$-$C)$_{\rm A}$}  & \colhead{(O$-$C)$_{\rm B}$} & \colhead{Orbital} \\
\colhead{\hbox{~~(2,400,000$+$)~~}} & \colhead{(\kms)} &
\colhead{(\kms)} & \colhead{(\kms)} & \colhead{(\kms)} & \colhead{Phase} }
\startdata
  45546.727\dotfill &   $-$19.53  & $+$14.51  &    $-$0.08 &   $+$3.08 &   0.6932 \\
  45563.682\dotfill &   $+$32.17  & $-$51.84  &    $+$0.71 &   $+$1.35 &   0.9890 \\
  45623.550\dotfill &   $+$30.44  & $-$52.12  &    $+$0.40 &   $-$0.74 &   0.0334 \\
  45688.410\dotfill &    $+$2.07  & $-$14.98  &    $-$0.02 &   $+$0.93 &   0.1649 \\
  45689.413\dotfill &    $-$1.49  & $-$10.75  &    $-$0.21 &   $+$0.88 &   0.1824 \\
  45694.419\dotfill &   $-$13.96  &  $+$5.34  &    $+$0.08 &   $+$0.78 &   0.2697 \\
  45695.426\dotfill &   $-$15.67  &  $+$8.57  &    $+$0.22 &   $+$1.66 &   0.2873 \\
  45721.434\dotfill &   $-$15.66  &  $+$8.75  &    $-$0.74 &   $+$3.07 &   0.7410 \\
  45742.957\dotfill &   $+$12.13  & $-$32.08  &    $-$0.63 &   $-$2.63 &   0.1165 \\
  45757.909\dotfill &   $-$22.44  & $+$16.21  &    $+$0.11 &   $+$0.84 &   0.3774 \\
  45887.911\dotfill &   $-$22.70  & $+$14.83  &    $-$0.11 &   $-$0.59 &   0.6453 \\
  45896.729\dotfill &    $-$7.20  &  $-$4.13  &    $-$0.17 &   $+$0.21 &   0.7991 \\
  45917.758\dotfill &    $+$1.72  & $-$14.66  &    $-$0.15 &   $+$0.97 &   0.1660 \\
  45935.641\dotfill &   $-$25.91  & $+$19.89  &    $-$0.21 &   $+$0.52 &   0.4780 \\
  45945.660\dotfill &   $-$22.10  & $+$13.39  &    $+$0.08 &   $-$1.51 &   0.6527 \\
  45949.647\dotfill &   $-$16.94  & $+$11.16  &    $-$0.06 &   $+$2.99 &   0.7223 \\
  45962.577\dotfill &   $+$25.87  & $-$46.00  &    $+$0.25 &   $-$0.22 &   0.9479 \\
  45981.527\dotfill &   $-$14.93  &  $+$5.53  &    $+$0.05 &   $-$0.23 &   0.2785 \\
  46005.529\dotfill &   $-$19.54  & $+$11.15  &    $-$0.41 &   $+$0.12 &   0.6972 \\
  46008.517\dotfill &   $-$14.41  &  $+$1.93  &    $-$0.44 &   $-$2.55 &   0.7493 \\
  46215.852\dotfill &   $-$22.06  & $+$13.87  &    $-$0.10 &   $-$0.74 &   0.3663 \\
  46246.856\dotfill &   $+$16.19  & $-$32.33  &    $+$0.08 &   $+$1.38 &   0.9072 \\
  46274.870\dotfill &   $-$24.09  & $+$14.82  &    $-$0.65 &   $-$1.67 &   0.3959 \\
  46310.591\dotfill &   $+$31.08  & $-$53.54  &    $-$0.33 &   $-$0.41 &   0.0191 \\
  46489.937\dotfill &    $+$6.06  & $-$20.05  &    $+$0.42 &   $+$0.37 &   0.1479 \\
  46493.936\dotfill &    $-$7.15  &  $-$1.58  &    $+$0.05 &   $+$2.53 &   0.2176 \\
  48199.555\dotfill &   $+$29.89  & $-$50.92  &    $+$0.02 &   $+$0.25 &   0.9728 \\
  51257.899\dotfill &   $-$19.04  & $+$10.90  &    $+$0.30 &   $-$0.39 &   0.3268 \\
  51258.890\dotfill &   $-$20.53  & $+$12.46  &    $+$0.05 &   $-$0.40 &   0.3440 \\
  51260.898\dotfill &   $-$22.53  & $+$13.91  &    $+$0.11 &   $-$1.57 &   0.3791 \\
  51261.884\dotfill &   $-$23.54  & $+$14.28  &    $-$0.09 &   $-$2.23 &   0.3963 \\
  51263.902\dotfill &   $-$24.36  & $+$18.06  &    $+$0.37 &   $-$0.08 &   0.4315 \\
  51264.901\dotfill &   $-$25.27  & $+$18.55  &    $-$0.08 &   $-$0.17 &   0.4489 \\
  51267.884\dotfill &   $-$25.59  & $+$19.63  &    $+$0.30 &   $+$0.02 &   0.5009 \\
  51268.880\dotfill &   $-$25.97  & $+$19.16  &    $-$0.06 &   $-$0.48 &   0.5183 \\
  51269.877\dotfill &   $-$25.62  & $+$18.30  &    $+$0.21 &   $-$1.23 &   0.5357 \\
  51270.870\dotfill &   $-$25.37  & $+$17.62  &    $+$0.26 &   $-$1.66 &   0.5530 \\
  51273.886\dotfill &   $-$24.28  & $+$16.68  &    $+$0.05 &   $-$0.95 &   0.6057 \\
  51274.884\dotfill &   $-$23.33  & $+$16.42  &    $+$0.33 &   $-$0.35 &   0.6231 \\
  51277.857\dotfill &   $-$20.64  & $+$13.07  &    $+$0.15 &   $-$0.07 &   0.6749 \\
  51279.868\dotfill &   $-$17.78  & $+$11.49  &    $+$0.25 &   $+$1.86 &   0.7100 \\
  51293.838\dotfill &   $+$26.71  & $-$47.09  &    $-$0.07 &   $+$0.16 &   0.9537 \\
  51295.873\dotfill &   $+$31.48  & $-$52.45  &    $+$0.00 &   $+$0.76 &   0.9892 \\
  51296.859\dotfill &   $+$32.24  & $-$52.68  &    $+$0.32 &   $+$1.09 &   0.0064 \\
  51297.814\dotfill &   $+$31.41  & $-$51.70  &    $+$0.30 &   $+$1.05 &   0.0231 \\
  51298.869\dotfill &   $+$28.23  & $-$50.08  &    $-$0.70 &   $-$0.10 &   0.0415 \\
  51299.858\dotfill &   $+$26.18  & $-$45.17  &    $+$0.26 &   $+$0.99 &   0.0587 \\
  51300.861\dotfill &   $+$22.61  & $-$40.39  &    $+$0.39 &   $+$1.07 &   0.0762 \\
  51380.600\dotfill &   $-$25.37  & $+$20.08  &    $+$0.18 &   $+$0.91 &   0.4673 \\
  51413.642\dotfill &   $+$28.58  & $-$49.05  &    $+$0.00 &   $+$0.49 &   0.0437 \\
  51424.724\dotfill &   $-$10.20  &  $-$0.90  &    $-$0.19 &   $-$0.34 &   0.2371 \\
  51445.726\dotfill &   $-$24.67  & $+$18.91  &    $-$0.26 &   $+$1.18 &   0.6035 \\
  51464.487\dotfill &   $+$21.74  & $-$41.97  &    $-$0.10 &   $-$0.99 &   0.9308 \\
  51466.675\dotfill &   $+$29.24  & $-$51.29  &    $-$0.10 &   $-$0.79 &   0.9689 \\
  51468.667\dotfill &   $+$32.45  & $-$52.45  &    $+$0.51 &   $+$1.35 &   0.0037 \\
  51476.569\dotfill &    $+$7.09  & $-$21.63  &    $+$0.06 &   $+$0.54 &   0.1415 \\
  51521.513\dotfill &   $+$20.77  & $-$42.93  &    $+$0.15 &   $-$3.50 &   0.9256 \\
  51525.510\dotfill &   $+$32.26  & $-$53.03  &    $+$0.48 &   $+$0.57 &   0.9953 \\
  51530.440\dotfill &   $+$20.79  & $-$42.20  &    $-$0.28 &   $-$2.20 &   0.0813 \\
  51618.884\dotfill &   $-$23.54  & $+$18.30  &    $+$0.06 &   $+$1.60 &   0.6243 \\
  51640.847\dotfill &   $+$31.85  & $-$53.88  &    $-$0.06 &   $-$0.12 &   0.0074 \\
  51667.839\dotfill &   $-$25.55  & $+$18.42  &    $+$0.16 &   $-$0.95 &   0.4783 \\
  51682.802\dotfill &   $-$14.67  & $+$10.21  &    $+$0.44 &   $+$4.29 &   0.7393 \\
  51699.727\dotfill &   $+$29.26  & $-$51.44  &    $-$0.63 &   $-$0.24 &   0.0346 \\
  51712.696\dotfill &   $-$12.89  &  $+$4.55  &    $+$0.13 &   $+$1.28 &   0.2609 \\
  51788.531\dotfill &   $-$24.41  & $+$18.05  &    $+$0.59 &   $-$0.43 &   0.5838 \\
  51808.604\dotfill &   $+$22.50  & $-$42.91  &    $-$0.10 &   $-$0.97 &   0.9340 \\
  51815.609\dotfill &   $+$26.25  & $-$47.50  &    $-$0.16 &   $-$0.73 &   0.0562 \\
  51842.553\dotfill &   $-$26.18  & $+$17.75  &    $-$0.29 &   $-$1.85 &   0.5263 \\
  51866.503\dotfill &   $+$24.55  & $-$46.12  &    $-$0.28 &   $-$1.34 &   0.9441 \\
  51874.448\dotfill &   $+$20.28  & $-$39.25  &    $-$0.47 &   $+$0.35 &   0.0827 \\
  51921.441\dotfill &   $+$14.54  & $-$33.52  &    $-$0.39 &   $-$1.31 &   0.9025 \\
  52004.892\dotfill &   $-$21.96  & $+$13.64  &    $-$0.47 &   $-$0.38 &   0.3583 \\
\enddata
\end{deluxetable}

\clearpage

\begin{deluxetable}{lccc@{}cc}
\tabletypesize{\scriptsize}
\tablenum{2}
\tablecolumns{6}
\tablewidth{49pc}
\tablecaption{Spectroscopic orbital solutions for HD~195987.\label{tab:sb2}}
\rotate
\tablehead{
\colhead{\hfil~~~~~~~~~~~Element~~~~~~~~~~~~} & \colhead{This paper} & \colhead{Imbert (1980)} & \colhead{Latham et al. (1988)} & \colhead{Duquennoy \& Mayor (1991)}
& \colhead{Goldberg et al.\ (2002)}}
\startdata
\multicolumn{6}{l}{Adjusted quantities} \\
~~~$P$ (days)\dotfill                       &  57.32161~$\pm$~0.00034\phn      &  57.3210~$\pm$~0.0014\phn    &  57.325~$\pm$~0.009\phn      &  57.3240~$\pm$~0.0013\phn    & 57.3199~$\pm$~0.0065\phn \\
~~~$\gamma$ (\kms)\dotfill                  &  $-5.867$~$\pm$~0.038\phs        &  $-5.630$~$\pm$~0.075\phs    &  $-5.56$~$\pm$~0.12\phs      &  $-6.13$~$\pm$~0.07\phs      & $-$5.59~$\pm$~0.12\phs   \\
~~~$K_{\rm A}$ (\kms)\dotfill               &  28.944~$\pm$~0.046\phn          &  28.80~$\pm$~0.12\phn        &  28.89~$\pm$~0.20\phn        &  28.73~$\pm$~0.10\phn        & 29.16~$\pm$~0.19\phn     \\
~~~$K_{\rm B}$ (\kms)\dotfill               &  36.73~$\pm$~0.21\phn            &  \nodata                     &  \nodata                     &  \nodata                     & 34.6~$\pm$~1.2\phn       \\
~~~$e$\dotfill                              &  0.3103~$\pm$~0.0018             &  0.305~$\pm$~0.003           &  0.316~$\pm$~0.006           &  0.306~$\pm$~0.003           & 0.3083~$\pm$~0.0060      \\
~~~$\omega_{\rm A}$ (deg)\dotfill           &  357.03~$\pm$~0.35\phn\phn       &  356.8~$\pm$~0.7\phn\phn     &  355.9~$\pm$~0.8\phn\phn     &  356.8~$\pm$~0.7\phn\phn     & 357.1~$\pm$~1.1\phn\phn  \\
~~~$T$ (HJD$-$2,400,000)\tablenotemark{a}\dotfill            &  49404.825~$\pm$~0.045\phm{2222} &  49404.6~$\pm$~0.1\phm{2222} &  49404.8~$\pm$~0.3\phm{2222} &  49404.933~$\pm$~0.097\phm{2222}  & 49404.78~$\pm$~0.16\phm{2222} \\
\multicolumn{6}{l}{Derived quantities} \\
~~~$f(M)$ (M$_{\sun}$)\dotfill              &  \nodata                         &  0.122~$\pm$~0.002           &  0.1225~$\pm$~0.0026         &  0.1219~$\pm$~0.0017         & \nodata \\
~~~$M_{\rm A}\sin^3 i$ (M$_{\sun}$)\dotfill &  0.808~$\pm$~0.010               &  \nodata                     &  \nodata                     &  \nodata                     & 0.721~$\pm$~0.052  \\ 
~~~$M_{\rm B}\sin^3 i$ (M$_{\sun}$)\dotfill &  0.6369~$\pm$~0.0046             &  \nodata                     &  \nodata                     &  \nodata                     & 0.607~$\pm$~0.024  \\
~~~$q\equiv M_{\rm B}/M_{\rm A}$\dotfill    &  0.7881~$\pm$~0.0047             &  \nodata                     &  \nodata                     &  \nodata                     & 0.84~$\pm$~0.03    \\
~~~$a_{\rm A}\sin i$ (10$^6$ km)\dotfill    &  21.689~$\pm$~0.036\phn          &  21.6~$\pm$~0.1\phn          &  21.65~$\pm$~0.16\phn        &  21.57~$\pm$~0.10\phn        & \nodata \\
~~~$a_{\rm B}\sin i$ (10$^6$ km)\dotfill    &  27.52~$\pm$~0.16\phn            &  \nodata                     &  \nodata                     &  \nodata                     & \nodata \\
~~~$a \sin i$ (R$_{\sun}$)\dotfill          &  70.70~$\pm$~0.24\phn            &  \nodata                     &  \nodata                     &  \nodata                     & 68.7~$\pm$~1.3\phn \\
\multicolumn{6}{l}{Other quantities pertaining to the fit} \\
~~~$N_{\rm obs}$\dotfill                    &  73$+$73                         &  60                          &  31                          &  64                          & 28$+$28  \\
~~~Time span (days)\dotfill                 &  6458                            &  758                         &  1003                        &  4365                        & 2653 \\
~~~$\sigma_{\rm A}$ (\kms)\tablenotemark{b}\dotfill          &  0.30                            &  0.54                        &  0.68                        &  0.54                        & 0.63 \\
~~~$\sigma_{\rm B}$ (\kms)\tablenotemark{b}\dotfill          &  1.38                            &  \nodata                     &  \nodata                     &  \nodata                     & 4.51 \\
\enddata
\tablenotetext{a}{The time of periastron passage in the four solutions from the
literature has been shifted by an integer number of cycles to the epoch derived
in this paper, for comparison purposes.}
\tablenotetext{b}{Root mean square residual from the fit.}
\end{deluxetable}

\clearpage

\begin{deluxetable}{cccccc}
\tablenum{3}
\tablecolumns{6}
\tablewidth{39pc}
 \tablecaption{Summary of the relevant parameters of the calibration
objects adopted for our analysis of the PTI visibilities.}

\tablehead{
\colhead{} & \colhead{} & \colhead{$V$} & \colhead{$K$} & \colhead{$\Delta\phi$\tablenotemark{a}}  & \colhead{Adopted Model Diameter\tablenotemark{b}} \\
\colhead{~~~~Object Name~~~~}   & \colhead{Spectral Type} & \colhead{(mag)} & \colhead{(mag)} & \colhead{($\deg$)} & \colhead{(mas)}
}

\startdata
HD 195194\dotfill & \ion{G8}{3}  & 7.0  &  4.8  & 2.7  & 0.67 $\pm$ 0.10   \\
HD 200031\dotfill & \ion{G5}{3}  & 6.8  &  4.7  & 6.0  & 0.56 $\pm$ 0.05   \\
HD 177196\dotfill & \ion{A7}{5}  & 5.0  &  4.5  & 17   & 0.62 $\pm$ 0.10   \\
HD 185395\dotfill & \ion{F4}{5}  & 4.5  &  3.5  & 17   & 0.84 $\pm$ 0.08   \\
\enddata
\tablenotetext{a}{Angular separation from HD~195987.}
\tablenotetext{b}{The angular diameters were determined from effective
temperature and bolometric flux estimates based on archival broad-band
photometry, and visibility measurements with PTI.}

\label{tab:calibrators}

\end{deluxetable}

\clearpage

\begin{deluxetable}{ccccrrc}
\tablenum{4}
\tablewidth{26pc}
\tabletypesize{\scriptsize}
\tablecaption{Visibility measurements of HD~195987 in the $H$ band, and residuals from the final fit.\label{tab:visibH}}
\tablehead{
\colhead{HJD} & \colhead{} & \colhead{} &
\colhead{}  & \colhead{$u$} & \colhead{$v$} & \colhead{Orbital} \\
\colhead{\hbox{~~(2,400,000$+$)~~}} & \colhead{$V^2$} &
\colhead{$\sigma_{V^2}$} & \colhead{(O$-$C)$_{V^2}$} & \colhead{($10^6\lambda$)} & 
\colhead{($10^6\lambda$)} & \colhead{Phase} }
\startdata
  51745.813\dotfill &  0.322 &  0.135 & $-$0.123 & $-$30.683 & $-$57.824 & 0.8386 \\
  51745.815\dotfill &  0.407 &  0.127 & $-$0.050 & $-$30.198 & $-$58.251 & 0.8386 \\
  51745.827\dotfill &  0.639 &  0.259 & $+$0.123 & $-$27.951 & $-$59.560 & 0.8388 \\
  51745.828\dotfill &  0.608 &  0.174 & $+$0.083 & $-$27.624 & $-$59.703 & 0.8389 \\
  51745.894\dotfill &  1.051 &  0.339 & $+$0.166 & $-$12.316 & $-$65.245 & 0.8400 \\
  51745.901\dotfill &  0.920 &  0.272 & $+$0.006 & $-$10.291 & $-$65.561 & 0.8401 \\
  51745.941\dotfill &  1.025 &  0.414 & $+$0.067 &  $+$0.135 & $-$66.600 & 0.8408 \\
  52103.920\dotfill &  0.743 &  0.231 & $+$0.015 & $-$44.026 & $-$26.884 & 0.0859 \\
  52103.922\dotfill &  0.740 &  0.031 & $-$0.019 & $-$43.768 & $-$27.217 & 0.0859 \\
  52103.946\dotfill &  1.054 &  0.341 & $+$0.080 & $-$39.373 & $-$31.461 & 0.0864 \\
  52175.716\dotfill &  0.584 &  0.037 & $+$0.001 & $-$45.253 & $-$25.448 & 0.3384 \\
  52175.719\dotfill &  0.638 &  0.025 & $-$0.048 & $-$44.871 & $-$25.984 & 0.3385 \\
  52175.729\dotfill &  0.911 &  0.084 & $-$0.039 & $-$43.101 & $-$27.937 & 0.3386 \\
  52175.740\dotfill &  0.956 &  0.050 & $+$0.123 & $-$41.183 & $-$29.784 & 0.3388 \\
  52175.742\dotfill &  0.846 &  0.047 & $+$0.073 & $-$40.779 & $-$30.131 & 0.3389 \\
  52175.775\dotfill &  0.481 &  0.038 & $-$0.014 & $-$33.720 & $-$35.322 & 0.3394 \\
  52175.777\dotfill &  0.499 &  0.033 & $-$0.077 & $-$33.055 & $-$35.692 & 0.3395 \\
  52175.788\dotfill &  0.854 &  0.101 & $+$0.004 & $-$30.364 & $-$37.135 & 0.3397 \\
  52175.799\dotfill &  0.905 &  0.094 & $+$0.029 & $-$27.430 & $-$38.497 & 0.3399 \\
  52177.609\dotfill &  0.980 &  0.063 & $+$0.049 & $-$50.012 &  $-$4.245 & 0.3714 \\
  52177.616\dotfill &  0.977 &  0.088 & $+$0.032 & $-$50.413 &  $-$5.798 & 0.3716 \\
  52177.662\dotfill &  0.789 &  0.066 & $+$0.034 & $-$49.975 & $-$15.690 & 0.3724 \\
  52177.671\dotfill &  0.977 &  0.132 & $+$0.008 & $-$49.397 & $-$17.432 & 0.3725 \\
  52177.710\dotfill &  0.534 &  0.106 & $+$0.072 & $-$45.059 & $-$25.216 & 0.3732 \\
  52177.712\dotfill &  0.662 &  0.086 & $+$0.109 & $-$44.757 & $-$25.646 & 0.3732 \\
  52177.730\dotfill &  0.944 &  0.194 & $+$0.012 & $-$41.940 & $-$28.921 & 0.3735 \\
  52177.738\dotfill &  0.651 &  0.139 & $-$0.025 & $-$40.397 & $-$30.397 & 0.3737 \\
  52177.740\dotfill &  0.643 &  0.089 & $+$0.045 & $-$39.995 & $-$30.719 & 0.3737 \\
  52179.617\dotfill &  0.972 &  0.059 & $+$0.028 & $-$50.720 &  $-$7.199 & 0.4065 \\
  52179.625\dotfill &  0.978 &  0.061 & $+$0.075 & $-$50.815 &  $-$8.877 & 0.4066 \\
  52179.662\dotfill &  0.692 &  0.072 & $+$0.045 & $-$49.836 & $-$16.701 & 0.4072 \\
  52179.669\dotfill &  0.942 &  0.082 & $+$0.019 & $-$49.322 & $-$18.258 & 0.4074 \\
  52179.704\dotfill &  0.373 &  0.059 & $+$0.027 & $-$45.322 & $-$25.308 & 0.4080 \\
  52179.706\dotfill &  0.446 &  0.027 & $+$0.036 & $-$45.064 & $-$25.641 & 0.4080 \\
  52179.713\dotfill &  0.802 &  0.077 & $+$0.061 & $-$43.940 & $-$27.019 & 0.4081 \\
  52179.731\dotfill &  0.736 &  0.068 & $-$0.029 & $-$40.884 & $-$30.179 & 0.4085 \\
  52179.739\dotfill &  0.415 &  0.058 & $-$0.026 & $-$39.393 & $-$31.485 & 0.4086 \\
\enddata
\end{deluxetable}

\clearpage

\begin{deluxetable}{ccccrrc}
\tablenum{5}
\tablewidth{26pc}
\tabletypesize{\scriptsize}
\tablecaption{Visibility measurements of HD~195987 in the $K$ band, and residuals from the final fit.\label{tab:visibK}}
\tablehead{
\colhead{HJD} & \colhead{} & \colhead{} &
\colhead{}  & \colhead{$u$} & \colhead{$v$} & \colhead{Orbital} \\
\colhead{\hbox{~~(2,400,000$+$)~~}} & \colhead{$V^2$} &
\colhead{$\sigma_{V^2}$} & \colhead{(O$-$C)$_{V^2}$} & \colhead{($10^6\lambda$)} & 
\colhead{($10^6\lambda$)} & \colhead{Phase} }
\startdata
  51353.925\dotfill &  0.749 &  0.042 & $+$0.021 & $-$16.722 & $-$46.365 & 0.0020 \\
  51353.949\dotfill &  0.945 &  0.044 & $+$0.015 & $-$12.573 & $-$47.823 & 0.0024 \\
  51353.974\dotfill &  0.792 &  0.040 & $-$0.034 &  $-$7.839 & $-$48.933 & 0.0028 \\
  51353.996\dotfill &  0.572 &  0.036 & $-$0.032 &  $-$3.398 & $-$49.354 & 0.0032 \\
  51354.912\dotfill &  0.360 &  0.035 & $-$0.048 & $-$18.491 & $-$45.828 & 0.0192 \\
  51354.935\dotfill &  0.723 &  0.070 & $-$0.023 & $-$14.629 & $-$47.393 & 0.0196 \\
  51354.958\dotfill &  0.892 &  0.100 & $-$0.039 & $-$10.290 & $-$48.625 & 0.0200 \\
  51354.999\dotfill &  0.667 &  0.069 & $-$0.016 &  $-$2.352 & $-$49.666 & 0.0207 \\
  51355.888\dotfill &  0.331 &  0.020 & $+$0.017 & $-$22.144 & $-$44.552 & 0.0362 \\
  51355.935\dotfill &  0.531 &  0.027 & $+$0.015 & $-$14.339 & $-$48.212 & 0.0370 \\
  51356.918\dotfill &  0.290 &  0.027 & $-$0.005 & $-$16.691 & $-$46.600 & 0.0542 \\
  51356.962\dotfill &  0.400 &  0.040 & $-$0.006 &  $-$8.602 & $-$48.910 & 0.0549 \\
  51356.985\dotfill &  0.674 &  0.053 & $-$0.009 &  $-$3.942 & $-$49.566 & 0.0553 \\
  51360.887\dotfill &  0.693 &  0.044 & $+$0.010 & $-$19.901 & $-$45.257 & 0.1234 \\
  51360.937\dotfill &  0.331 &  0.021 & $+$0.016 & $-$11.361 & $-$48.578 & 0.1243 \\
  51360.938\dotfill &  0.312 &  0.010 & $+$0.005 & $-$11.031 & $-$48.563 & 0.1243 \\
  51375.901\dotfill &  0.369 &  0.059 & $+$0.031 & $-$10.375 & $-$48.550 & 0.3853 \\
  51385.791\dotfill &  0.340 &  0.025 & $-$0.067 & $-$24.229 & $-$43.148 & 0.5579 \\
  51385.833\dotfill &  0.628 &  0.021 & $+$0.025 & $-$17.894 & $-$46.942 & 0.5586 \\
  51386.788\dotfill &  0.265 &  0.015 & $-$0.035 & $-$24.378 & $-$43.274 & 0.5753 \\
  51386.815\dotfill &  0.860 &  0.031 & $+$0.000 & $-$20.403 & $-$45.756 & 0.5757 \\
  51386.848\dotfill &  0.420 &  0.021 & $+$0.029 & $-$14.703 & $-$48.179 & 0.5763 \\
  51399.766\dotfill &  0.427 &  0.016 & $-$0.021 & $-$22.243 & $-$43.927 & 0.8017 \\
  51399.802\dotfill &  0.181 &  0.022 & $-$0.026 & $-$16.355 & $-$46.763 & 0.8023 \\
  51399.843\dotfill &  0.350 &  0.022 & $-$0.012 &  $-$8.860 & $-$49.032 & 0.8030 \\
  51401.762\dotfill &  0.689 &  0.040 & $-$0.042 & $-$21.834 & $-$43.935 & 0.8365 \\
  51401.803\dotfill &  0.914 &  0.037 & $+$0.020 & $-$15.229 & $-$47.152 & 0.8372 \\
  51401.841\dotfill &  0.980 &  0.043 & $+$0.004 &  $-$8.129 & $-$49.167 & 0.8379 \\
  51415.769\dotfill &  0.873 &  0.037 & $-$0.044 & $-$14.464 & $-$47.350 & 0.0809 \\
  51415.789\dotfill &  0.766 &  0.051 & $+$0.025 & $-$10.851 & $-$48.303 & 0.0812 \\
  51416.744\dotfill &  0.511 &  0.032 & $+$0.095 & $-$18.247 & $-$45.574 & 0.0979 \\
  51416.772\dotfill &  0.795 &  0.052 & $+$0.068 & $-$13.292 & $-$47.424 & 0.0984 \\
  51472.637\dotfill &  0.386 &  0.088 & $+$0.024 & $-$10.119 & $-$48.863 & 0.0729 \\
  51694.947\dotfill &  0.269 &  0.024 & $+$0.023 & $-$23.303 & $-$42.350 & 0.9512 \\
  51694.962\dotfill &  0.246 &  0.026 & $+$0.015 & $-$21.226 & $-$43.806 & 0.9515 \\
  51694.973\dotfill &  0.299 &  0.032 & $+$0.016 & $-$19.566 & $-$44.658 & 0.9517 \\
  51694.992\dotfill &  0.456 &  0.068 & $-$0.011 & $-$16.527 & $-$46.106 & 0.9520 \\
  51695.002\dotfill &  0.766 &  0.073 & $+$0.182 & $-$14.756 & $-$46.764 & 0.9522 \\
  51707.915\dotfill &  0.581 &  0.074 & $+$0.134 & $-$22.996 & $-$42.913 & 0.1774 \\
  51707.929\dotfill &  0.469 &  0.066 & $+$0.009 & $-$20.956 & $-$44.202 & 0.1777 \\
  51707.956\dotfill &  0.632 &  0.080 & $+$0.162 & $-$16.666 & $-$46.282 & 0.1782 \\
  51707.971\dotfill &  0.415 &  0.082 & $-$0.052 & $-$14.123 & $-$47.192 & 0.1784 \\
  51707.975\dotfill &  0.483 &  0.143 & $+$0.015 & $-$13.324 & $-$47.500 & 0.1785 \\
  51707.989\dotfill &  0.430 &  0.142 & $-$0.030 & $-$10.823 & $-$48.181 & 0.1787 \\
  51707.992\dotfill &  0.493 &  0.118 & $+$0.039 & $-$10.127 & $-$48.263 & 0.1788 \\
  51708.003\dotfill &  0.525 &  0.136 & $+$0.080 &  $-$8.120 & $-$48.712 & 0.1790 \\
  51716.888\dotfill &  0.276 &  0.045 & $-$0.046 & $-$23.348 & $-$42.595 & 0.3340 \\
  51716.898\dotfill &  0.286 &  0.029 & $+$0.020 & $-$21.859 & $-$43.446 & 0.3342 \\
  51716.953\dotfill &  1.015 &  0.075 & $+$0.135 & $-$12.895 & $-$47.520 & 0.3351 \\
  51716.963\dotfill &  0.935 &  0.101 & $+$0.078 & $-$11.014 & $-$48.043 & 0.3353 \\
  51716.990\dotfill &  0.565 &  0.059 & $-$0.052 &  $-$5.850 & $-$48.819 & 0.3358 \\
  51717.001\dotfill &  0.570 &  0.065 & $+$0.065 &  $-$3.743 & $-$49.051 & 0.3359 \\
  51723.876\dotfill &  0.874 &  0.067 & $+$0.025 & $-$22.044 & $-$42.796 & 0.4559 \\
  51723.890\dotfill &  0.873 &  0.075 & $+$0.044 & $-$20.086 & $-$43.970 & 0.4561 \\
  51723.919\dotfill &  0.284 &  0.046 & $-$0.044 & $-$15.387 & $-$46.184 & 0.4566 \\
  51723.933\dotfill &  0.426 &  0.061 & $+$0.024 & $-$12.929 & $-$46.946 & 0.4569 \\
  51723.949\dotfill &  0.715 &  0.066 & $+$0.045 &  $-$9.883 & $-$47.728 & 0.4572 \\
  51723.963\dotfill &  0.866 &  0.080 & $+$0.055 &  $-$7.236 & $-$48.294 & 0.4574 \\
  51723.974\dotfill &  0.822 &  0.224 & $+$0.014 &  $-$5.106 & $-$48.406 & 0.4576 \\
  51723.988\dotfill &  0.686 &  0.089 & $+$0.001 &  $-$2.333 & $-$48.729 & 0.4579 \\
  51723.996\dotfill &  0.710 &  0.227 & $+$0.105 &  $-$0.918 & $-$48.809 & 0.4580 \\
  51741.797\dotfill &  0.976 &  0.055 & $+$0.048 & $-$36.915 &  $-$2.364 & 0.7685 \\
  51741.804\dotfill &  0.735 &  0.090 & $-$0.153 & $-$37.236 &  $-$3.540 & 0.7687 \\
  51741.902\dotfill &  0.397 &  0.019 & $-$0.024 & $-$33.607 & $-$18.647 & 0.7704 \\
  51741.909\dotfill &  0.559 &  0.036 & $+$0.019 & $-$32.877 & $-$19.660 & 0.7705 \\
  51741.944\dotfill &  1.082 &  0.038 & $+$0.094 & $-$28.023 & $-$24.126 & 0.7711 \\
  51741.951\dotfill &  1.042 &  0.041 & $+$0.066 & $-$26.960 & $-$24.913 & 0.7712 \\
  51744.796\dotfill &  0.254 &  0.024 & $-$0.011 & $-$37.043 &  $-$3.596 & 0.8209 \\
  51744.803\dotfill &  0.235 &  0.032 & $-$0.021 & $-$37.251 &  $-$4.670 & 0.8210 \\
  51744.851\dotfill &  0.200 &  0.035 & $-$0.003 & $-$36.705 & $-$12.146 & 0.8218 \\
  51744.858\dotfill &  0.191 &  0.020 & $-$0.013 & $-$36.328 & $-$13.234 & 0.8219 \\
  51744.928\dotfill &  0.484 &  0.047 & $-$0.024 & $-$29.065 & $-$23.065 & 0.8232 \\
  51744.935\dotfill &  0.542 &  0.077 & $-$0.023 & $-$28.034 & $-$23.957 & 0.8233 \\
  51746.822\dotfill &  0.206 &  0.045 & $+$0.000 & $-$21.122 & $-$44.293 & 0.8562 \\
  51746.824\dotfill &  0.226 &  0.049 & $+$0.019 & $-$20.782 & $-$44.390 & 0.8562 \\
  51746.836\dotfill &  0.253 &  0.044 & $+$0.042 & $-$19.017 & $-$45.391 & 0.8564 \\
  51746.839\dotfill &  0.287 &  0.086 & $+$0.076 & $-$18.522 & $-$45.707 & 0.8565 \\
  51746.893\dotfill &  0.301 &  0.023 & $+$0.047 &  $-$8.869 & $-$48.761 & 0.8574 \\
  51746.895\dotfill &  0.300 &  0.099 & $+$0.043 &  $-$8.427 & $-$48.856 & 0.8575 \\
  51752.793\dotfill &  0.460 &  0.032 & $-$0.008 & $-$37.534 &  $-$6.427 & 0.9604 \\
  51752.799\dotfill &  0.596 &  0.040 & $+$0.017 & $-$37.575 &  $-$7.511 & 0.9605 \\
  51752.806\dotfill &  0.701 &  0.058 & $+$0.007 & $-$37.545 &  $-$8.587 & 0.9606 \\
  51752.829\dotfill &  0.955 &  0.075 & $-$0.015 & $-$36.897 & $-$12.119 & 0.9610 \\
  51752.835\dotfill &  0.999 &  0.082 & $+$0.011 & $-$36.571 & $-$13.156 & 0.9611 \\
  51752.842\dotfill &  0.944 &  0.066 & $-$0.024 & $-$36.237 & $-$14.146 & 0.9612 \\
  51752.872\dotfill &  0.468 &  0.048 & $-$0.040 & $-$33.585 & $-$18.607 & 0.9617 \\
  51791.675\dotfill &  0.927 &  0.084 & $-$0.007 & $-$24.303 & $-$41.631 & 0.6387 \\
  51791.683\dotfill &  0.925 &  0.068 & $+$0.082 & $-$23.256 & $-$42.516 & 0.6388 \\
  51794.671\dotfill &  0.282 &  0.050 & $+$0.019 & $-$23.622 & $-$41.847 & 0.6909 \\
  51794.714\dotfill &  0.995 &  0.096 & $+$0.064 & $-$17.173 & $-$45.574 & 0.6917 \\
  51794.780\dotfill &  0.420 &  0.051 & $+$0.047 &  $-$5.164 & $-$48.662 & 0.6928 \\
  51831.634\dotfill &  0.981 &  0.282 & $+$0.117 & $-$13.669 & $-$47.076 & 0.3358 \\
  51831.644\dotfill &  0.957 &  0.364 & $+$0.077 & $-$11.832 & $-$47.527 & 0.3359 \\
  51832.645\dotfill &  0.618 &  0.038 & $-$0.016 & $-$11.295 & $-$48.013 & 0.3534 \\
  51832.655\dotfill &  0.513 &  0.057 & $+$0.026 &  $-$9.402 & $-$48.524 & 0.3536 \\
  51832.687\dotfill &  0.327 &  0.032 & $-$0.006 &  $-$3.219 & $-$49.407 & 0.3541 \\
  51832.703\dotfill &  0.388 &  0.051 & $-$0.017 &  $+$0.028 & $-$49.456 & 0.3544 \\
  51832.741\dotfill &  0.613 &  0.079 & $-$0.006 &  $+$7.522 & $-$48.791 & 0.3551 \\
  51832.749\dotfill &  0.747 &  0.185 & $+$0.080 &  $+$9.092 & $-$48.683 & 0.3552 \\
  51833.587\dotfill &  0.779 &  0.073 & $+$0.054 & $-$20.747 & $-$44.399 & 0.3698 \\
  51833.594\dotfill &  0.797 &  0.076 & $-$0.030 & $-$19.686 & $-$44.961 & 0.3700 \\
  51833.628\dotfill &  0.542 &  0.064 & $-$0.062 & $-$13.791 & $-$47.277 & 0.3706 \\
  51833.636\dotfill &  0.370 &  0.090 & $-$0.100 & $-$12.457 & $-$47.827 & 0.3707 \\
  51833.671\dotfill &  0.404 &  0.065 & $+$0.020 &  $-$5.797 & $-$49.071 & 0.3713 \\
  51833.678\dotfill &  0.456 &  0.055 & $+$0.005 &  $-$4.458 & $-$49.294 & 0.3714 \\
  51833.713\dotfill &  0.875 &  0.229 & $+$0.136 &  $+$2.534 & $-$49.312 & 0.3720 \\
  51833.720\dotfill &  0.959 &  0.244 & $+$0.184 &  $+$3.949 & $-$49.290 & 0.3722 \\
  51850.576\dotfill &  0.299 &  0.206 & $+$0.092 & $-$35.890 & $-$14.261 & 0.6662 \\
  51850.587\dotfill &  0.333 &  0.143 & $-$0.066 & $-$35.123 & $-$16.014 & 0.6664 \\
  51850.589\dotfill &  0.774 &  0.457 & $+$0.331 & $-$34.953 & $-$16.216 & 0.6664 \\
  51850.627\dotfill &  0.842 &  0.286 & $+$0.130 & $-$30.666 & $-$21.528 & 0.6671 \\
  51850.636\dotfill &  0.425 &  0.144 & $+$0.020 & $-$29.413 & $-$22.756 & 0.6673 \\
  51851.630\dotfill &  0.586 &  0.128 & $-$0.095 & $-$29.867 & $-$22.233 & 0.6846 \\
  51851.632\dotfill &  0.588 &  0.039 & $-$0.025 & $-$29.662 & $-$22.568 & 0.6846 \\
  52120.799\dotfill &  0.244 &  0.015 & $-$0.020 & $-$37.790 &  $-$8.720 & 0.3804 \\
  52120.806\dotfill &  0.435 &  0.028 & $+$0.017 & $-$37.598 &  $-$9.746 & 0.3805 \\
  52120.821\dotfill &  0.933 &  0.039 & $+$0.063 & $-$37.127 & $-$12.097 & 0.3807 \\
  52120.827\dotfill &  1.037 &  0.034 & $+$0.058 & $-$36.800 & $-$13.153 & 0.3809 \\
  52120.850\dotfill &  0.517 &  0.030 & $-$0.009 & $-$35.175 & $-$16.534 & 0.3812 \\
  52154.730\dotfill &  0.700 &  0.112 & $+$0.016 & $-$17.204 & $-$46.179 & 0.9723 \\
  52154.753\dotfill &  1.056 &  0.075 & $+$0.153 & $-$13.188 & $-$47.569 & 0.9727 \\
  52154.776\dotfill &  1.025 &  0.099 & $+$0.114 &  $-$8.897 & $-$48.621 & 0.9731 \\
  52156.711\dotfill &  0.447 &  0.017 & $+$0.023 & $-$19.236 & $-$44.927 & 0.0069 \\
  52156.741\dotfill &  0.924 &  0.032 & $+$0.086 & $-$14.316 & $-$46.944 & 0.0074 \\
  52156.767\dotfill &  0.949 &  0.025 & $+$0.030 &  $-$9.421 & $-$48.279 & 0.0078 \\
  52156.788\dotfill &  0.715 &  0.025 & $-$0.046 &  $-$5.397 & $-$48.945 & 0.0082 \\
  52180.651\dotfill &  0.373 &  0.026 & $-$0.017 & $-$37.374 & $-$11.148 & 0.4245 \\
  52180.658\dotfill &  0.657 &  0.027 & $+$0.021 & $-$37.059 & $-$12.252 & 0.4246 \\
  52180.671\dotfill &  1.034 &  0.036 & $+$0.052 & $-$36.298 & $-$14.354 & 0.4249 \\
  52180.678\dotfill &  0.955 &  0.031 & $+$0.013 & $-$35.826 & $-$15.419 & 0.4250 \\
  52180.692\dotfill &  0.458 &  0.034 & $-$0.028 & $-$34.620 & $-$17.493 & 0.4252 \\
  52180.699\dotfill &  0.280 &  0.018 & $-$0.007 & $-$33.960 & $-$18.397 & 0.4253 \\
  52180.713\dotfill &  0.310 &  0.017 & $+$0.004 & $-$32.398 & $-$20.408 & 0.4256 \\
  52180.720\dotfill &  0.543 &  0.021 & $+$0.018 & $-$31.499 & $-$21.269 & 0.4257 \\
\enddata
\end{deluxetable}

\clearpage

\begin{deluxetable}{lcccc}
\tabletypesize{\scriptsize}
\tablenum{6}
\tablecolumns{5}
\tablewidth{40pc}
\tablecaption{Orbital parameters for HD~195987.\label{tab:combsol}}
\rotate
\tablehead{\colhead{~~~~~~~~~~~Parameter~~~~~~~~~~~} & \colhead{HIPPARCOS\tablenotemark{a}} & \colhead{$V^2$ Only} & \colhead{RV Only} & \colhead{Full Fit}}
\startdata
\multicolumn{5}{l}{Adjusted quantities} \\
~~~$P$ (days)\dotfill                      & 57.3240           & 57.3298~$\pm$~0.0035\phn  & 57.32161~$\pm$~0.00034\phn     & 57.32178~$\pm$~0.00029\phn  \\
~~~$a$ (mas)\tablenotemark{b}\dotfill      & 5.24~$\pm$~0.66   & 15.368~$\pm$~0.028\phn    & \nodata                        & 15.378~$\pm$~0.027\phn      \\
~~~$\gamma$ ($\kms$)\dotfill               & \nodata           & \nodata                   & $-$5.867~$\pm$~0.038\phm{$-$}  & $-$5.841~$\pm$~0.037\phm{$-$}      \\
~~~$K_A$ ($\kms$)\dotfill                  & \nodata           & \nodata                   & 28.944~$\pm$~0.046\phn         & 28.929~$\pm$~0.046\phn      \\
~~~$K_B$ ($\kms$)\dotfill                  & \nodata           & \nodata                   & 36.73~$\pm$~0.21\phn           & 36.72~$\pm$~0.21\phn        \\
~~~$e$\dotfill                             & 0.3060            & 0.30740~$\pm$~0.00067     & 0.3103~$\pm$~0.0018            & 0.30626~$\pm$~0.00057   \\
~~~$i$ (deg)\dotfill                       & 89.50~$\pm$~8.43\phn  & 99.379~$\pm$~0.088\phn & \nodata                       & 99.364~$\pm$~0.080\phn      \\
~~~$\omega_A$ (deg)\dotfill                & 356.8             & 358.89~$\pm$~0.53\phn\phn  & 357.03~$\pm$~0.35\phn\phn     & 357.40~$\pm$~0.29\phn\phn       \\
~~~$\Omega$ (deg)\tablenotemark{c}\dotfill & 327.66~$\pm$~7.56\phn\phn & 335.061~$\pm$~0.082\phn\phn  & \nodata             & 334.960~$\pm$~0.070\phn\phn     \\
~~~$T$ (HJD$-$2,400,000)\dotfill             & 43327.589\tablenotemark{d}         & 51354.000~$\pm$~0.069\phm{2222} & 49404.825~$\pm$~0.045\phm{2222}  & 51353.813~$\pm$~0.038\phm{2222}   \\
~~~$\Delta K_{\rm CIT}$ (mag)\dotfill      & \nodata           & 1.063~$\pm$~0.031         & \nodata                        & 1.056~$\pm$~0.013     \\
~~~$\Delta H_{\rm CIT}$ (mag)\dotfill      & \nodata           & 1.18~$\pm$~0.16           & \nodata                        & 1.154~$\pm$~0.065       \\
\multicolumn{5}{l}{Derived quantities} \\
~~~$\pi$ (mas)\tablenotemark{e}\dotfill    & 45.30~$\pm$~0.46\phn  &  \nodata  &  \nodata  &  46.08~$\pm$~0.27\phn \\
\multicolumn{5}{l}{Other quantities pertaining to the fit} \\
~~~$N_{\rm obs}$ (RV)\dotfill      &  \nodata  &  \nodata  &  73$+$73 & 73$+$73 \\
~~~$N_{\rm obs}$ ($V^2$)\dotfill   &  \nodata  & \phn37$+$134 & \nodata & \phn37$+$134 \\
~~~$\sigma_{V^2}$ ($H$/$K$)\tablenotemark{f}\dotfill     &  \nodata  & 0.0599~/~0.0613 & \nodata & 0.0592~/~0.0617  \\
~~~$\sigma_{\rm RV}$ (A/B, \kms)\tablenotemark{f}\dotfill     &  \nodata  & \nodata & 0.30~/~1.38  & 0.31~/~1.41\\
\enddata
\tablenotetext{a}{The elements $P$, $e$, $\omega_A$, and $T$ in this solution were adopted from Duquennoy \& Mayor (1991) and held fixed.}
\tablenotetext{b}{For the HIPPARCOS solution this is the semimajor axis of the photocenter ($\alpha$) rather than the relative semimajor axis.}
\tablenotetext{c}{Position angle of the ascending node, i.e., the node at which the secondary is receding.}
\tablenotetext{d}{This value listed in the HIPPARCOS catalog is erroneous; it should be 43328.589 (cf.\ Duquennoy \& Mayor 1991).}
\tablenotetext{e}{The determination from the HIPPARCOS data is our revised value (see \S4), and the full-fit determination is the orbital parallax.}
\tablenotetext{f}{Root mean square residual from the fit.}
\end{deluxetable}

\clearpage

\begin{deluxetable}{lcc}
\tablenum{7}
\tablecolumns{2}
\tablewidth{24pc}
\tablecaption{Physical parameters of HD~195987.\label{tab:physics}}
\tablehead{
\colhead{~~~~~~Parameter~~~~~~} & \colhead{Primary} & \colhead{Secondary}
}

\startdata
Mass (M$_{\sun}$)\dotfill             &  0.844~$\pm$~0.018   & 0.6650~$\pm$~0.0079 \\
$T_{\rm eff}$ (K)\dotfill             &  5200~$\pm$~100\phn  & 4200~$\pm$~200\phn  \\
$\pi_{\rm orb}$ (mas)\dotfill         &  \multicolumn{2}{c}{46.08~$\pm$~0.27\phn}      \\
Dist (pc)\dotfill                     &  \multicolumn{2}{c}{21.70~$\pm$~0.13\phn}      \\
$M_V$~~(mag)\dotfill                  &  5.511~$\pm$~0.028   & 7.91~$\pm$~0.19     \\
$M_H$\tablenotemark{a}~~(mag)\dotfill &  3.679~$\pm$~0.037   & 4.835~$\pm$~0.059 \\
$M_K$\tablenotemark{a}~~(mag)\dotfill &  3.646~$\pm$~0.033   & 4.702~$\pm$~0.034   \\
$V\!-\!K$\tablenotemark{a}~~(mag)\dotfill & 1.865~$\pm$~0.039 & 3.21~$\pm$~0.19     \\
\enddata
\tablenotetext{a}{In the CIT system.}
\end{deluxetable}

\end{document}